\documentclass[aps,twocolumn,superscriptaddress,amsmath,longbibliography]{revtex4-1}
\usepackage{amsmath}
\usepackage{amsfonts}
\usepackage{amssymb}
\usepackage{braket}
\usepackage{xcolor}
\usepackage{graphicx}
\usepackage[section]{placeins}

\begin{document}

\title{Distinct Critical Behaviors from the Same State in Quantum Spin \\ and Population Dynamics Perspectives}

\author{C. L. Baldwin}
\affiliation{National Institute of Standards and Technology, Gaithersburg, MD 20899, USA}
\affiliation{Joint Quantum Institute, University of Maryland, College Park, MD 20742, USA}

\author{S. Shivam}
\affiliation{Department of Physics, Princeton University, Princeton, New Jersey 08544, USA}

\author{S. L. Sondhi}
\affiliation{Department of Physics, Princeton University, Princeton, New Jersey 08544, USA}

\author{M. Kardar}
\affiliation{Department of Physics, Massachusetts Institute of Technology, Cambridge, Massachusetts 02139, USA}

\begin{abstract}

There is a deep connection between the ground states of transverse-field spin systems and the late-time distributions of evolving viral populations -- within simple models, both are obtained from the  principal eigenvector of the same matrix.
However, that vector is the wavefunction amplitude in the quantum spin model, whereas it is the probability itself in the population model.
We show that this seemingly minor difference has significant consequences: phase transitions which are discontinuous in the spin system become continuous when viewed through the population perspective, and transitions which are continuous become governed by new critical exponents.
We introduce a more general class of models which encompasses both cases, and that can be solved exactly in a mean-field limit.
Numerical results are also presented for a number of one-dimensional 
chains with power-law interactions.
We see that well-worn spin models of quantum statistical mechanics can contain unexpected new physics and insights when treated as population-dynamical models and beyond, motivating further studies.

\end{abstract}

\maketitle

\tableofcontents

\section{Introduction}

\begin{figure}[t]
\centering
\includegraphics[width=1.0\columnwidth]{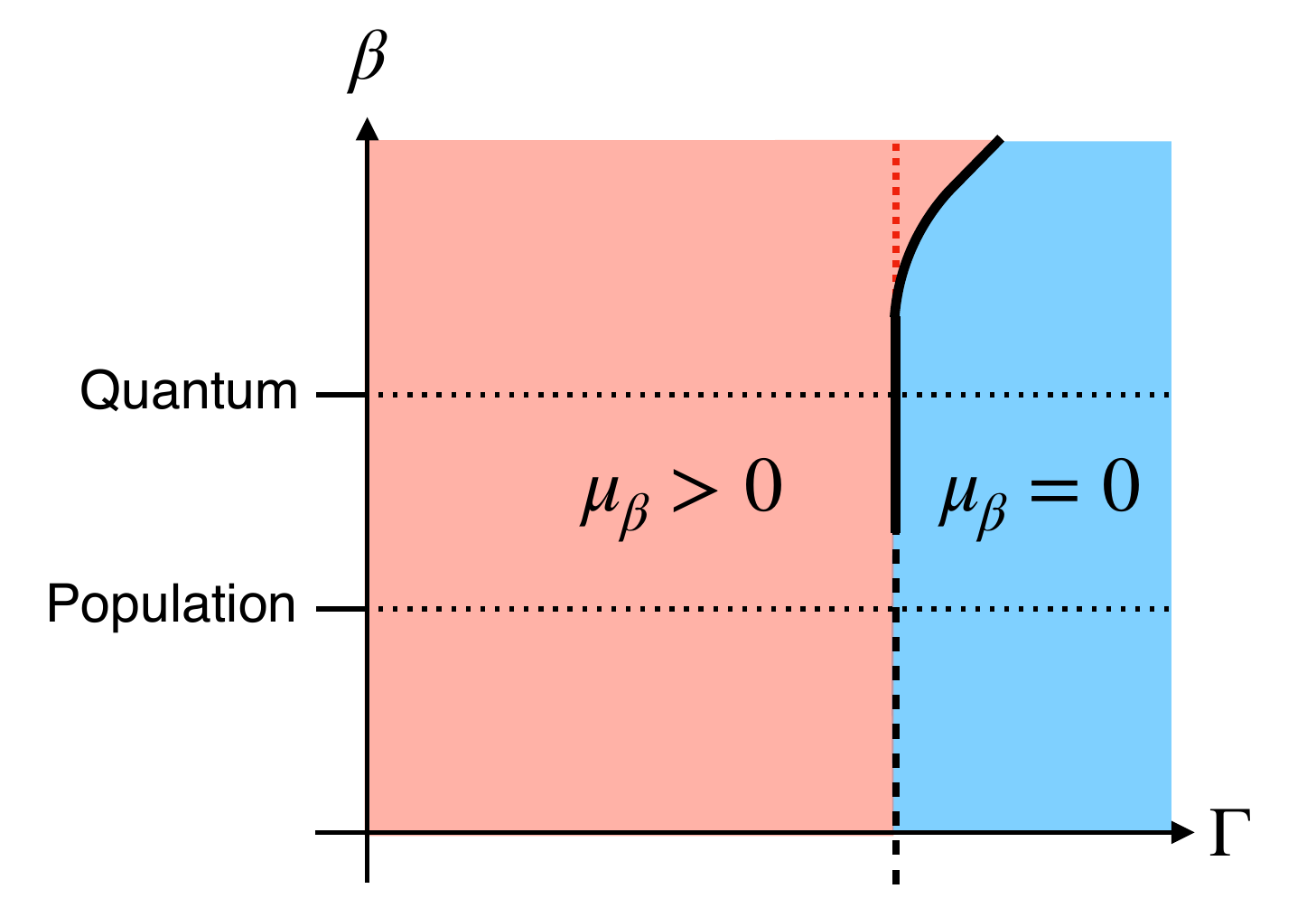}
\caption{Sketch of a typical phase diagram in the $\Gamma$-$\beta$ plane (see the discussion following Eq.~\eqref{eq:DNA_Hamiltonian}). Red shading indicates the ordered phase and blue indicates disordered. The black dashed line indicates a continuous transition and solid indicates discontinuous, and the red dashed line is a transition between two ordered phases. The quantum model corresponds to the line $\beta = 2$, and the population dynamics model corresponds to $\beta = 1$ (both dotted).}
\label{fig:phase_diagram}
\end{figure}

In a somewhat simplified perspective, the evolution of viral populations is governed by two competing processes: mutation of the genetic code upon reproduction, and natural selection due to differences in the corresponding reproduction rates.
Mutations destroy the information contained in the genetic sequence and lead to a wider variety of sequences in the population (known as a quasi-species cloud), whereas selection promotes those sequences which give the fastest reproduction rates at the expense of slower members.
The quasi-species population collapses if the rate of mutations is too large, 
suggesting a sharp transition-- an ``error catastrophe''-- in the number of mutations per virus~\cite{Eigen1971Selforganization,Wilke2005Quasispecies}.
It has motivated the treatment of RNA viruses such as HIV through hypermutation: increasing the average mutation rate in the viral population so as to drastically reduce the proportion of viable members~\cite{Crotty2001RNA,Zhang2003Cytidine,Anderson2004Viral,GrandePerez2005Suppression,Hart2015Error,Gupta2015Scaling,Shivam2020Studying}.

A particularly simple model for mutation-selection dynamics is to represent genetic sequences by chains of Ising spins: $\sigma \equiv \{ \sigma_i \} _{i=1}^N$, where $\sigma_i = -1$ indicates a mutation on site $i$ and $\sigma_i = 1$ indicates no mutation (called ``wild-type'').
The wild-type state on site $i$ changes to the mutated state at rate $\Gamma_i^+$, and the mutated state reverts to wild-type at rate $\Gamma_i^-$.
Each sequence $\sigma$ reproduces at a certain rate $F(\sigma)$, called the fitness function.
Natural selection is captured by the fact that different $\sigma$ have different values of $F(\sigma)$.

A useful measure of the relative strength of mutation versus selection is the surplus $\mu_1$, defined as the average value throughout the population of $N^{-1} \sum_i \sigma_i$, i.e., the number of wild-type sites minus the number of mutated sites.
Clearly smaller $\Gamma_i^{\pm}$ and steeper $F(\sigma)$ favor $\mu_1 \approx 1$ (assuming the wild-type state has highest fitness), while larger $\Gamma_i^{\pm}$ and shallower $F(\sigma)$ favor $\mu_1 < 1$.

Note that the competition between mutation and selection is analogous to the competition between the two terms of a transverse-field Ising model: in an effort to lower the total energy, the transverse field encourages spin flips whereas the spin-spin interactions bias the system towards specific configurations having lower interaction energy.
The surplus is analogous to the magnetization of the Ising model, and an error catastrophe is a phase transition in the usual sense of statistical mechanics, i.e., non-analyticity of an observable~\footnote{Many works use a slightly different definition of error catastrophe, namely when the fraction of wild-type states in the population becomes zero. We use the definition involving average surplus because it is more natural from a statistical-physics perspective. See as well Ref.~\cite{Franz1997Error}.}.

In formulating the above considerations mathematically, we see that the relationship to a transverse-field Ising model is much more than an analogy.
The state of the (quasi-)population at time $t$ is indicated by the number of members having each possible sequence, $\mathcal{N}(\sigma, t)$.
Denote by $L_i \sigma$ the sequence differing from $\sigma$ only in the value at site $i$.
The time evolution of the population is then given by the equations
\begin{equation} \label{eq:population_equations_explicit}
\begin{aligned}
\frac{\textrm{d}}{\textrm{d}t} \mathcal{N}(\sigma, t) =& \, F(\sigma) \mathcal{N}(\sigma, t) \\
& + \sum_i \big( \Gamma_i^{-\sigma_i} \mathcal{N}(L_i \sigma, t) - \Gamma_i^{\sigma_i} \mathcal{N}(\sigma, t) \big) .
\end{aligned}
\end{equation}
The first term is the change due to reproduction, and the second term is that due to mutation.
We write Eq.~\eqref{eq:population_equations_explicit} more compactly by denoting $\mathcal{N}(\sigma, t)$ as a vector $| \mathcal{N}(t) \rangle$ in the $2^N$-dimensional Hilbert space having basis states $| \sigma \rangle$ (i.e., so that $\langle \sigma | \mathcal{N}(t) \rangle = \mathcal{N}(\sigma, t)$).
Evolution according to Eq.~\eqref{eq:population_equations_explicit} is then cast in the matrix form
\begin{equation} \label{eq:DNA_master_equation}
\frac{\textrm{d}}{\textrm{d}t} | \mathcal{N}(t) \rangle = -H | \mathcal{N}(t) \rangle,
\end{equation}
\begin{equation} \label{eq:DNA_Hamiltonian}
\begin{aligned}
H \equiv & -F \big( \hat{\sigma}^z \big) + \sum_i \left( \frac{\Gamma_i^+ + \Gamma_i^-}{2} + \frac{\Gamma_i^+ - \Gamma_i^-}{2} \hat{\sigma}_i^z \right) \\
& - \sum_i \big( \Gamma_i^+ \hat{\sigma}_i^+ + \Gamma_i^- \hat{\sigma}_i^- \big) ,
\end{aligned}
\end{equation}
with $\hat{\sigma}$ being the standard Pauli operators.

Equation~\eqref{eq:DNA_Hamiltonian} is quite literally the Hamiltonian of a transverse-field Ising model (albeit non-Hermitian unless $\Gamma_i^+ = \Gamma_i^-$), and Eq.~\eqref{eq:DNA_master_equation} can equally be seen as the imaginary-time Schrodinger equation.
In particular, as $t \rightarrow \infty$, the state $| \mathcal{N}(t) \rangle$ approaches the ground state of the Hamiltonian.
The steady-state value of the surplus in the population is seen to be a ground state property of the associated Ising Hamiltonian, analogous to the longitudinal magnetization, and any error catastrophe corresponds to a quantum phase transition.

Despite this deep connection between the error catastrophe and a quantum phase transition, the purpose of the present paper is to show that the \textit{nature} of the transition is often qualitatively different when viewed through the surplus rather than the magnetization.
One common means of classifying phase transitions is by the non-analyticity of an order parameter, e.g., continuous versus discontinuous.
We shall show that the surplus can go to zero continuously even when the magnetization is discontinuous, and can have novel critical exponents at continuous transitions.
As will become clear, these differences stem from one detail which was glossed over in the above discussion: the weight $\langle \sigma | \mathcal{N}(t) \rangle$ (once normalized) is the probability of observing configuration $\sigma$ when sampling randomly from the population, whereas if $| \mathcal{N}(t) \rangle$ were a quantum state it would be the square root of the probability.

The equivalence between equations for population dynamics and quantum Ising models is not new~\cite{Baake1997Ising,Wagner1998Ising}.
There have been observations that the corresponding order parameters can have different continuity properties~\cite{Tarazona1992Error,Hermisson2001Four} (although some specific models have turned out to be misleading~\cite{Franz1997Error}), and these observations have been explained in a purely mathematical sense~\cite{Baake1998Quantum,Hermisson2002Mutation}.
Yet in our opinion, such explanations, valuable as they are, do not give much physical intuition and risk making the correspondence between the two fields seem less powerful than it is.
Our aim in this paper is to study the problem using the techniques and terminology of quantum statistical physics, with the hope of encouraging further investigation of population-dynamical models among the condensed-matter physics community.

Furthermore, we place these results in the context of a larger family of models, taking the probabilities to be the weights $\langle \sigma | \mathcal{N}(t) \rangle$ raised to an arbitrary power $\beta$ ($\beta = 1$ corresponds to the population dynamics model and $\beta = 2$ corresponds to standard quantum mechanics).
This reveals intricate $\Gamma$-$\beta$ phase diagrams, one example of which is sketched in Fig.~\ref{fig:phase_diagram}.
We show that the nature of the phase transition in $\Gamma$ can depend on $\beta$ in a variety of ways, with the overall trends that the transition becomes continuous at lower $\beta$ and the critical field begins increasing at larger $\beta$.
The full significance of this non-trivial $\beta$-dependence remains to be discovered, but it is already useful in elucidating our results on surplus and magnetization.

In Sec.~\ref{sec:exact_solutions}, we present the analytical treatment of symmetric models, i.e., models in which the fitness function depends solely on the total magnetization.
Although idealized, they often serve as valuable toy systems among both the statistical physics and population genetics communities~\cite{Hermisson2002Mutation,Peliti2002Quasispecies,Saakian2004Solvable,Bapst2012On,Zhao2014Three}.
We show that the models commonly used to demonstrate discontinuous magnetic phase transitions generically have a continuous surplus.
In Sec.~\ref{sec:numerics}, we then present numerical results demonstrating the same phenomena in non-symmetric models.
Although finite-size effects prevent any quantitative conclusions, we do find evidence that the surplus often has distinct critical exponents at continuous phase transitions.
Finally, in Sec.~\ref{sec:discussion}, we highlight the relationship to a third topic which provides further understanding for these results: the critical behavior of free surfaces as compared to bulk systems.

\section{Exact solution of symmetric models} \label{sec:exact_solutions}

Symmetric Hamiltonians constitute a large family of models for which we can determine the ground state analytically, at least to leading order in large $N$.
By symmetric, we mean any fitness function $F(\sigma)$ which depends only on the total spin-$z$ $M(\sigma) \equiv \sum_i \sigma_i$.
An example is
\begin{equation} \label{eq:infinite_range_Hamiltonian}
F_0(\sigma) = \frac{1}{N} \sum_{i,j} \sigma_i \sigma_j = \frac{1}{N} M(\sigma)^2,
\end{equation}
which can equivalently be thought of as an Ising model with infinite-range interactions.
More generally, we write
\begin{equation} \label{eq:symmetric_fitness_normalization}
F(\sigma) = N f \left( \frac{M(\sigma)}{N} \right) ,
\end{equation}
where the factors of $N$ are included simply for convenience in what follows.
Furthermore, to make closer contact with the models used in statistical physics, we shall restrict ourselves to Hermitian Hamiltonians ($\Gamma_i^+ = \Gamma_i^-$).

\subsection{Definitions \& notation}

Taking $| \mathcal{N} \rangle$ to be the ground state of Eq.~\eqref{eq:DNA_Hamiltonian}, we denote $\langle \sigma | \mathcal{N} \rangle$ by $C_{\sigma}$.
Note that by the Perron-Frobenius theorem, $C_{\sigma} \geq 0$ for all $\sigma$.
The symmetry of the Hamiltonian ensures that the eigenstates have definite total angular momentum, and we shall focus on the subspace of maximal angular momentum $N$.
In this subspace, $C_{\sigma}$ is identical for all configurations having the same $M(\sigma)$.
We shall henceforth write $C_M$, where $M \in \{ -N, -N+2, \cdots, N \}$.

We will find that $C_M$ is, to leading order, exponentially small in $N$.
In particular,
\begin{equation} \label{eq:wavefunction_amplitude_scaling}
C_M \sim e^{-N \alpha(m)},
\end{equation}
for some smooth function $\alpha$ of $m \equiv M/N$.

By definition, the magnetization density of $| \mathcal{N} \rangle$ when viewed as a quantum state is
\begin{equation} \label{eq:magnetization_definition}
\mu_2 \equiv \frac{1}{N} \frac{\sum_{\sigma} M(\sigma) C_{M(\sigma)}^2}{\sum_{\sigma} C_{M(\sigma)}^2}.
\end{equation}
Correspondingly, the surplus density of $| \mathcal{N} \rangle$ when viewed as a population is
\begin{equation} \label{eq:surplus_definition}
\mu_1 \equiv \frac{1}{N} \frac{\sum_{\sigma} M(\sigma) C_{M(\sigma)}}{\sum_{\sigma} C_{M(\sigma)}}.
\end{equation}
Note that we can write
\begin{equation} \label{eq:observable_averages}
\mu_2 = \frac{1}{N} \sum_M M \big| \Psi(M) \big| ^2, \quad \mu_1 = \frac{1}{N} \sum_M M P(M),
\end{equation}
where
\begin{equation} \label{eq:observable_distributions}
\begin{aligned}
\big| \Psi(M) \big| ^2 =& \frac{\sum_{\sigma} \delta_{M(\sigma), M} C_{M(\sigma)}^2}{\sum_{\sigma} C_{M(\sigma)}^2} \propto \binom{N}{\frac{N+M}{2}} C_M^2, \\
P(M) =& \frac{\sum_{\sigma} \delta_{M(\sigma), M} C_{M(\sigma)}}{\sum_{\sigma} C_{M(\sigma)}} \propto \binom{N}{\frac{N+M}{2}} C_M,
\end{aligned}
\end{equation}
i.e., $| \Psi(M) |^2$ and $P(M)$ are the probability distributions for the magnetization and surplus respectively.

At large $N$, the binomial coefficient can be approximated as ($m \equiv M/N$)
\begin{equation} \label{eq:binomial_Stirling_approximation}
\binom{N}{\frac{N+M}{2}} \sim e^{N h(m)},
\end{equation}
where
\begin{equation} \label{eq:binomial_entropy}
h(m) = -\frac{1+m}{2} \log{\frac{1+m}{2}} - \frac{1-m}{2} \log{\frac{1-m}{2}}.
\end{equation}
Using Eq.~\eqref{eq:wavefunction_amplitude_scaling}, we have that
\begin{equation} \label{eq:observable_distributions_scaling}
\big| \Psi(M) \big| ^2 \propto e^{N \big( h(m) - 2 \alpha(m) \big) }, \quad P(M) \propto e^{N \big( h(m) - \alpha(m) \big) }.
\end{equation}
To leading order as $N\to\infty$,
\begin{equation} \label{eq:observable_averages_scaling}
\begin{aligned}
\mu_2 \sim & \; \textrm{argmax} \big[ h(m) - 2 \alpha(m) \big] , \\
\mu_1 \sim & \; \textrm{argmax} \big[ h(m) - \alpha(m) \big] .
\end{aligned}
\end{equation}
where argmax denotes the value of $m$ for which the argument is maximum.
Generalizing slightly, we can define an entire family of distributions
\begin{equation} \label{eq:generalized_surplus_distribution}
P_{\beta}(M) = \frac{\sum_{\sigma} \delta_{M(\sigma), M} C_{M(\sigma)}^{\beta}}{\sum_{\sigma} C_{M(\sigma)}^{\beta}} \propto e^{N s_{\beta}(m)},
\end{equation}
where $s_{\beta}(m) = h(m) - \beta \alpha(m)$, for an arbitrary positive real number $\beta$.
The generalized magnetization $\mu_{\beta}$ is defined as the expectation value with respect to $P_{\beta}(m)$ (hence the notation $\mu_2$ for magnetization and $\mu_1$ for surplus).
Although we do not have a physical interpretation for $\mu_{\beta}$ at arbitrary $\beta$, it will be useful to consider $\beta$ as a tunable parameter.

In the calculations that follow, it will be easier to work directly with $\Psi(M)$ rather than $C_M$, thus we give the exponent a name:
\begin{equation} \label{eq:magnetization_wavefunction_exponent}
\frac{1}{N} \log{\Psi(M)} \equiv \phi(m) = \frac{1}{2} h(m) - \alpha(m).
\end{equation}

To summarize, in the following section we shall calculate $\phi(m)$, then determine $s_{\beta}(m)$ via
\begin{equation} \label{eq:generalized_surplus_entropy}
s_{\beta}(m) = \left( 1 - \frac{\beta}{2} \right) h(m) + \beta \phi(m),
\end{equation}
and finally $\mu_{\beta}$ via
\begin{equation} \label{eq:generalized_surplus_average}
\mu_{\beta} = \textrm{argmax} \big[ s_{\beta}(m) \big] .
\end{equation}

\begin{figure}[t]
\centering
\includegraphics[width=1.0\columnwidth]{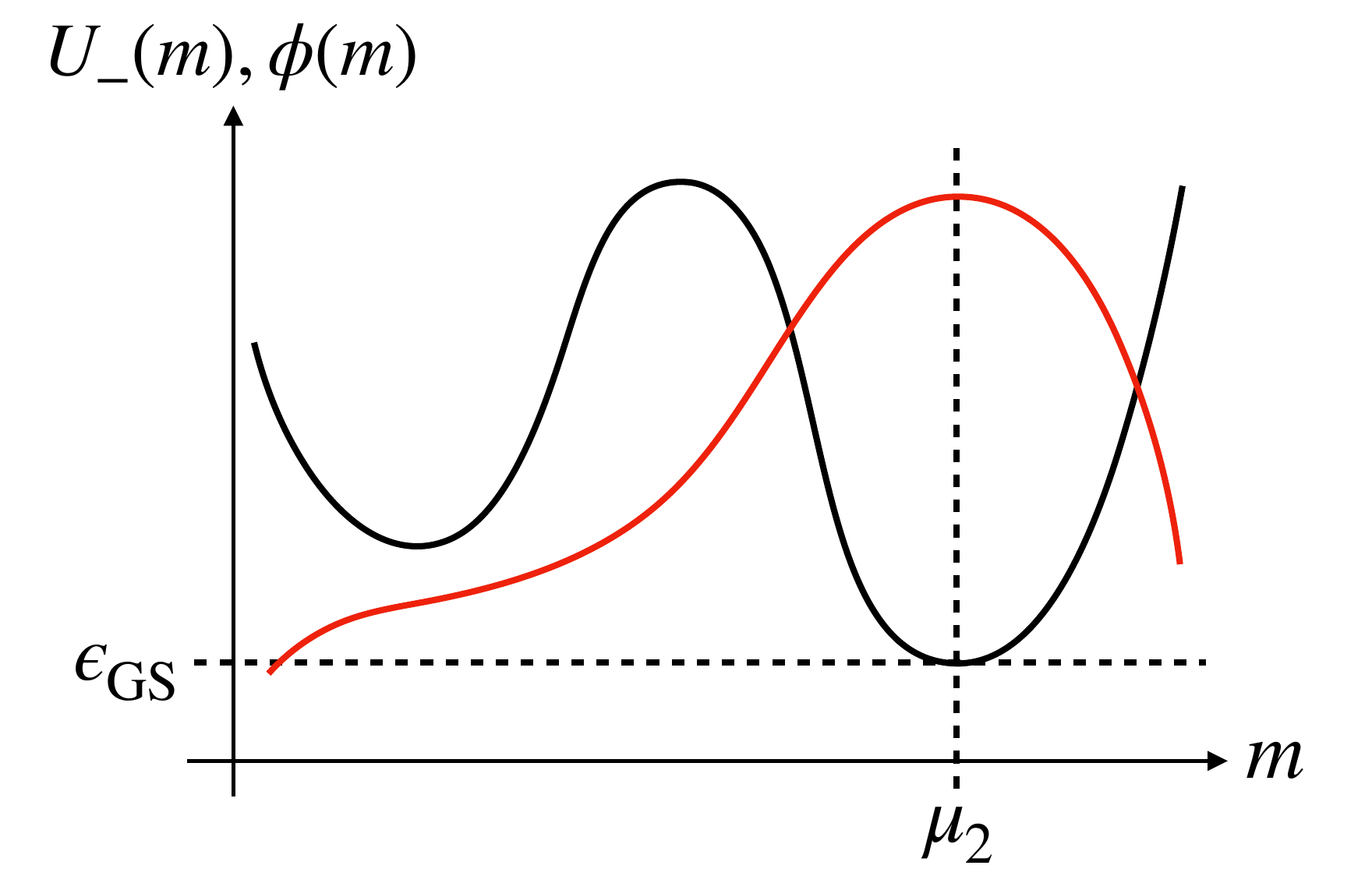}
\caption{Sketch of an example potential $U_-(m)$ (solid black line), the ground state energy density $\epsilon_{\textrm{GS}}$ and average magnetization $\mu_2$, and the resulting wavefunction exponent $\phi(m)$ (red line).}
\label{fig:potential_cartoon}
\end{figure}

\subsection{Large-$N$ analysis}

The eigenstates of $H$ in the subspace of maximal angular momentum can be determined analytically using the WKB method, which becomes exact in the $N \rightarrow \infty$ limit.
This technique, or equivalent formulations of it, has been applied successfully in both the quantum physics and population genetics fields~\cite{Garg1998Application,Hermisson2002Mutation,Saakian2007New,Bapst2012On}.

Noting that $\langle M | \mathcal{N} \rangle = \Psi(M)$ as defined above (where $| M \rangle$ is the basis state having total spin-$z$ $M$), the eigenvalue equation for $H$ can be written
\begin{equation} \label{eq:symmetric_Schrodinger_equation}
\begin{aligned}
E \Psi(M) =& -N f \left( \frac{M}{N} \right) \Psi(M) \\
& - \frac{\Gamma}{2} \sqrt{(N + M)(N - M + 2)} \Psi(M-2) \\
& - \frac{\Gamma}{2} \sqrt{(N - M)(N + M + 2)} \Psi(M+2).
\end{aligned}
\end{equation}
We write both $\log{\Psi(M)}$ and $E$ as series in $N$:
\begin{equation} \label{eq:wavefunction_large_N_expansion}
\Psi(M) = e^{N \phi(m) + \phi_1(m) + \frac{1}{N} \phi_2(m) + \cdots},
\end{equation}
\begin{equation} \label{eq:energy_large_N_expansion}
E = N \epsilon + \epsilon_1 + \frac{1}{N} \epsilon_2 + \cdots,
\end{equation}
then insert into Eq.~\eqref{eq:symmetric_Schrodinger_equation} and equate like powers of $N$ (while expanding terms like $\phi(m \pm \frac{2}{N})$ in Taylor series).
For our purposes, only the $O(N)$ equation will be needed.
It is
\begin{equation} \label{eq:WKB_zeroth_order_equation}
\epsilon = -f(m) - \Gamma \sqrt{1 - m^2} \cosh{\left( 2 \frac{\textrm{d}\phi}{\textrm{d}m} \right) }.
\end{equation}
Solving for $\textrm{d}\phi / \textrm{d}m$, we have
\begin{equation} \label{eq:WKB_leading_derivative}
\begin{gathered}
\frac{\textrm{d}\phi}{\textrm{d}m} = \frac{1}{2} \log{\Big( \kappa(m) \pm \sqrt{\kappa(m)^2 - 1} \Big) }, \\
\quad \kappa(m) \equiv \frac{-\epsilon - f(m)}{\Gamma \sqrt{1 - m^2}}.
\end{gathered}
\end{equation}

As discussed in Appendix~\ref{app:boundary_conditions}, the correct sign to use in Eq.~\eqref{eq:WKB_leading_derivative} is the plus sign near $m = -1$ and the minus sign near $m = 1$.
This requires that $|\kappa(m)|$ cross 1 at some intermediate value of $m$, so that $\textrm{d}\phi / \textrm{d}m$ is non-analytic there~\footnote{This is analogous to the situation in single-particle bound state problems, where states must have energies greater than the minimum of the potential in order to be normalizable.}.
The requirement that $|\kappa(m)| \leq 1$ for at least one point $m$ translates to a restriction on the allowed values of $\epsilon$: there must be a point $m$ at which
\begin{equation} \label{eq:WKB_spectrum_limits}
\begin{gathered}
U_-(m) \, \leq \, \epsilon \, \leq \, U_+(m), \\
U_{\pm}(m) \equiv -f(m) \pm \Gamma \sqrt{1 - m^2}.
\end{gathered}
\end{equation}
The ground state energy is the lowest allowed value:
\begin{equation} \label{eq:WKB_ground_state_energy}
\epsilon_{\textrm{GS}} = \textrm{min}_m \big[ U_-(m) \big] .
\end{equation}
These equations are best understood graphically, such as in Fig.~\ref{fig:potential_cartoon}.

Equation~\eqref{eq:WKB_spectrum_limits} further has a nice physical interpretation:
Consider a classical spin $\hat{s}$, by which we mean a unit vector in $\mathbb{R}^3$, with an energy function $H_{\textrm{cl}}(\hat{s})$ analogous to the original Hamiltonian:
\begin{equation} \label{eq:classical_analogue_energy}
H_{\textrm{cl}} \big( \hat{s} \big) = -f \big( s_z \big) - \Gamma s_x,
\end{equation}
where $s_x$ and $s_z$ are the projections along the $x$ and $z$ axes.
If $s_z$ is fixed to be $m$, then $s_x$ can take values between $-\sqrt{1 - m^2}$ and $\sqrt{1 - m^2}$.
$U_+(m)$ and $U_-(m)$ are precisely the maximum and minimum corresponding energies, and the lowest possible energy is found by minimizing $U_-(m)$, i.e., Eq.~\eqref{eq:WKB_ground_state_energy}.

The magnetization density of the ground state is correspondingly
\begin{equation} \label{eq:WKB_ground_state_magnetization}
\mu_2 = \textrm{argmin}_m \big[ U_-(m) \big] .
\end{equation}
This follows from having $\textrm{d}\phi / \textrm{d}m > 0$ for $m$ less than the argmin and $\textrm{d}\phi / \textrm{d}m < 0$ for $m$ greater than the argmin:
\begin{widetext}
\begin{equation} \label{eq:WKB_ground_state_wavefunction}
\frac{\textrm{d}\phi}{\textrm{d}m} = \begin{cases} \frac{1}{2} \log{\Big( \kappa(m) + \sqrt{\kappa(m)^2 - 1} \Big) }, \; \; & m \leq \textrm{argmin} [U_-] \\ \frac{1}{2} \log{\Big( \kappa(m) - \sqrt{\kappa(m)^2 - 1} \Big) }, \; \; & m \geq \textrm{argmin} [U_-] \end{cases}.
\end{equation}
\end{widetext}
Thus $\phi(m)$ is maximized at the argmin.
Since $\mu_2$ is the sum over $M$ of $M |\Psi(M)|^2$ and $\Psi(M)$ scales exponentially with $N$, the sum is dominated by where the exponent is maximal, giving Eq.~\eqref{eq:WKB_ground_state_magnetization}.
The situation is sketched in Fig.~\ref{fig:potential_cartoon}.

With this analysis in hand, we now calculate $\phi(m)$ and (through Eqs.~\eqref{eq:generalized_surplus_entropy} and~\eqref{eq:generalized_surplus_average}) $\mu_{\beta}$ for specific symmetric Hamiltonians.

\subsection{Results}

\begin{figure}[t]
\centering
\includegraphics[width=0.9\linewidth]{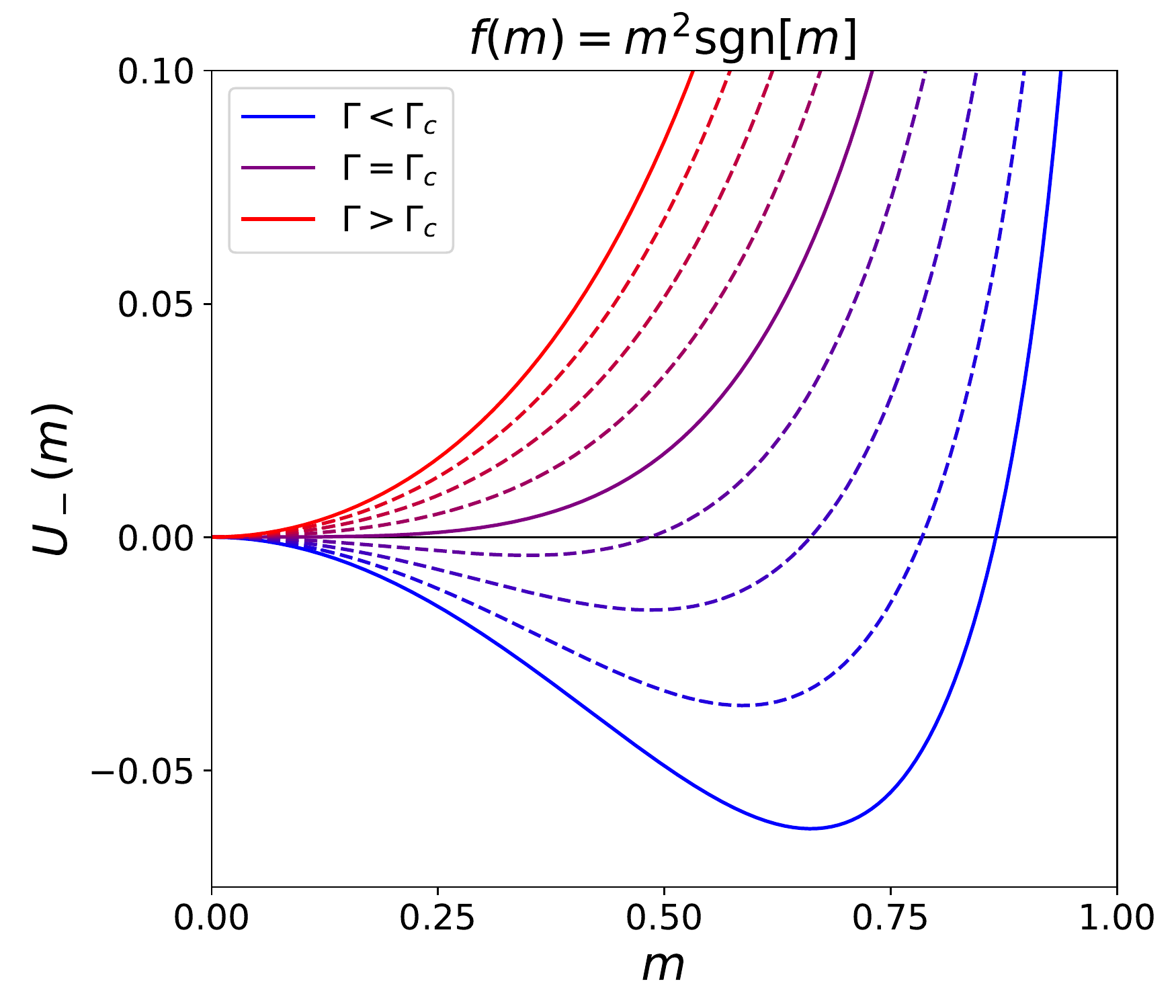}
\includegraphics[width=0.9\linewidth]{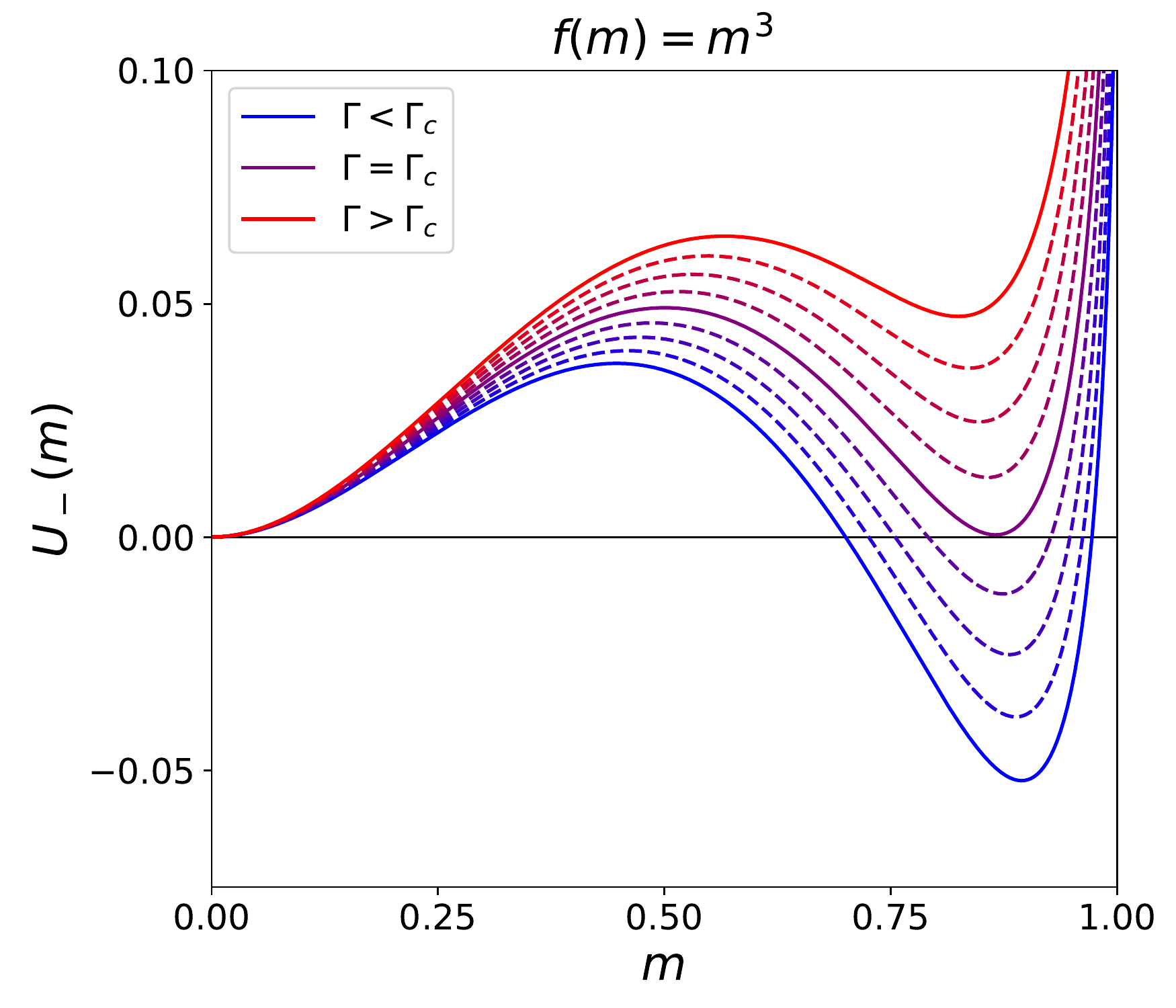}
\caption{Two examples of the potential $U_-(m)$ as a function of $m$, for various $\Gamma$ (increasing from blue curves to red). The fitness functions $f(m)$ are indicated, and the potential is given by Eq.~\eqref{eq:WKB_spectrum_limits}. For this figure, constants have been added to $U_-(m)$ so that $U_-(0) = 0$. (Top) A potential which gives a continuous transition. The $\Gamma$ values for the solid curves are 1.5, 2.0, 2.5 (blue to red). (Bottom) A potential which gives a discontinuous transition. $\Gamma$ values for the solid curves are 1.2, 1.3, 1.4.}
\label{fig:potential}
\end{figure}

\begin{figure}[t]
\centering
\includegraphics[width=0.9\linewidth]{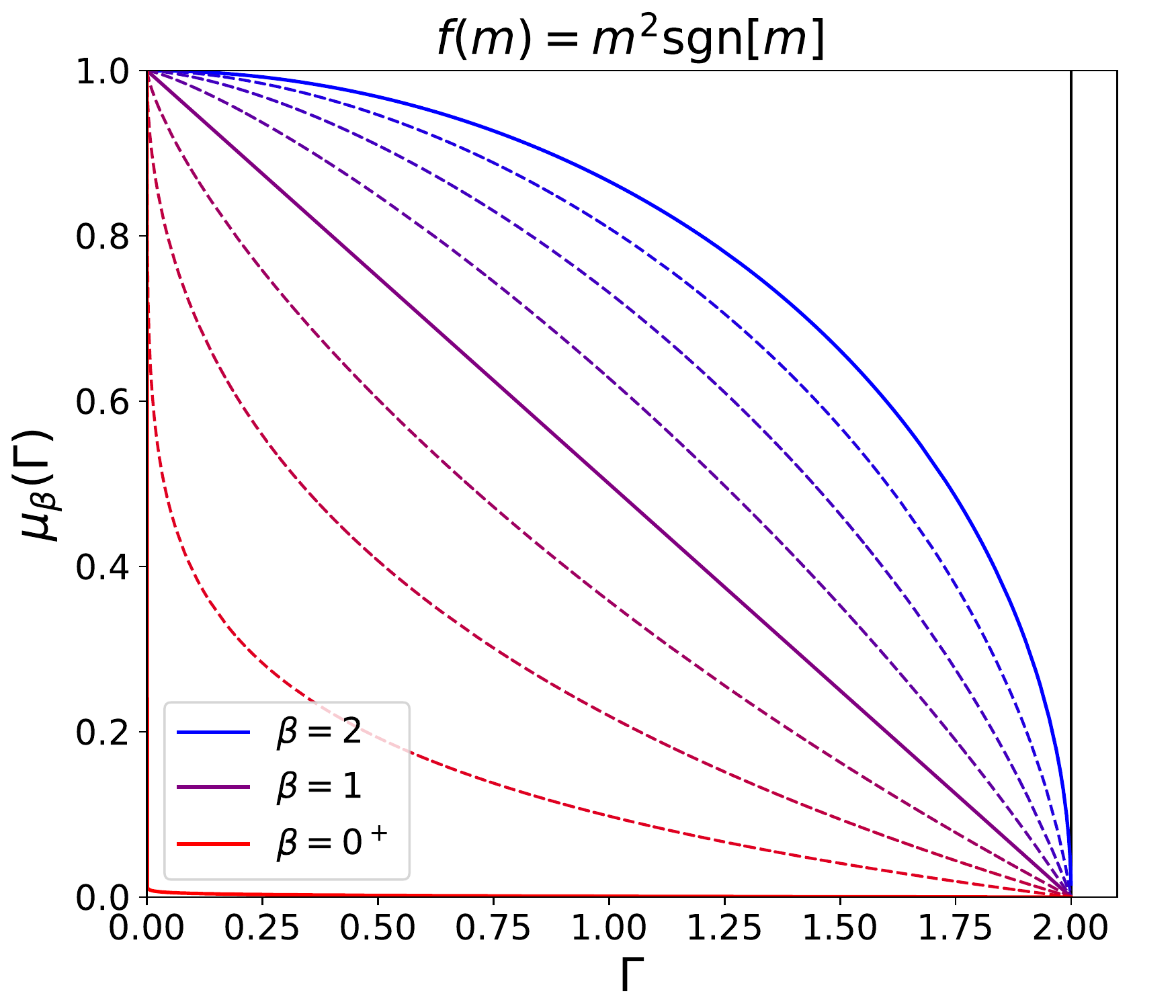}
\includegraphics[width=0.9\linewidth]{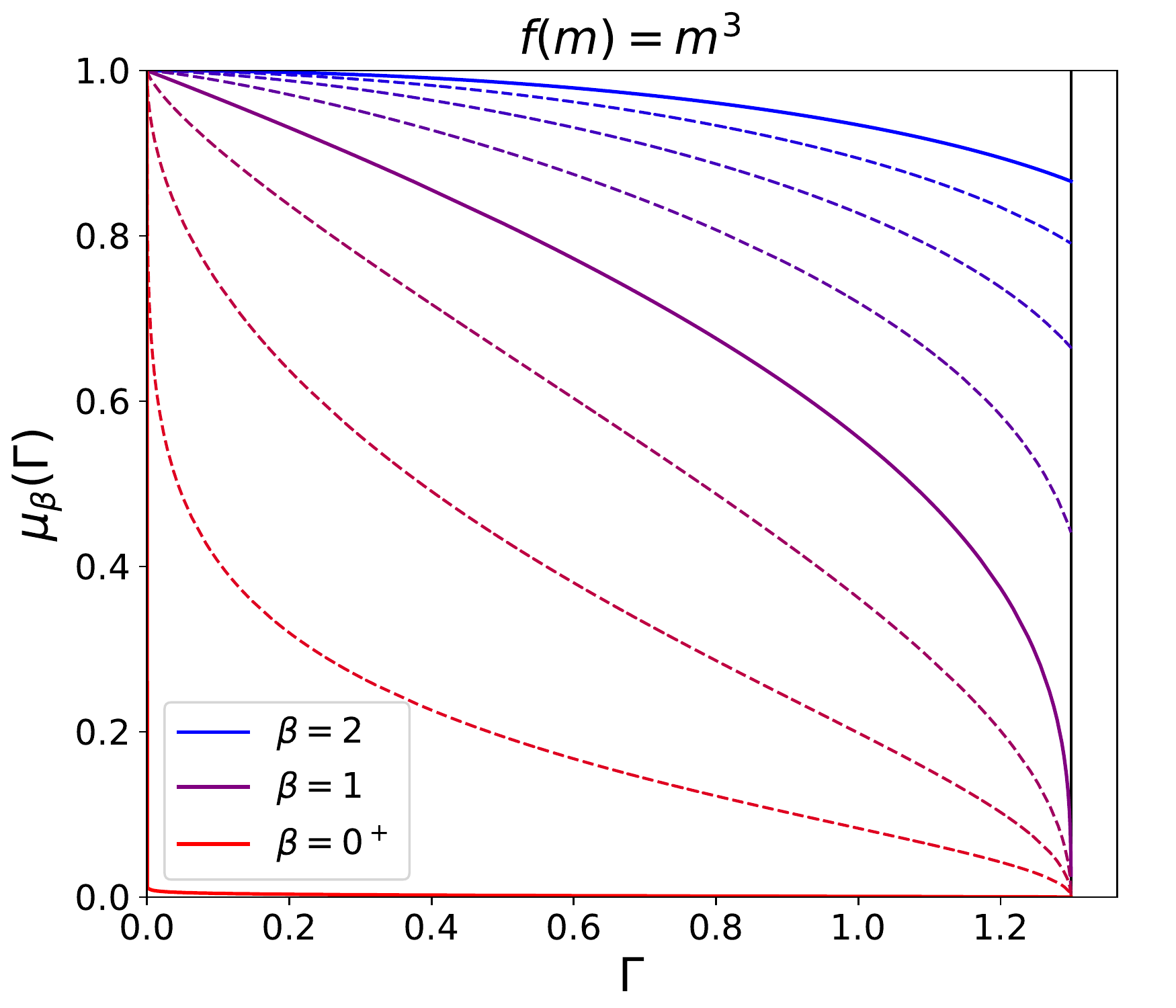}
\caption{Generalized magnetization $\mu_{\beta}$ as a function of $\Gamma$, for various $\beta$ (decreasing from blue to red) and the same fitness functions as in Fig.~\ref{fig:potential}. The vertical black lines indicate the values of $\Gamma_c$ ($\mu_{\beta}$ is identically 0 for $\Gamma > \Gamma_c$).}
\label{fig:general_mag}
\end{figure}

\begin{figure*}[t]
\centering
\includegraphics[width=0.9\textwidth]{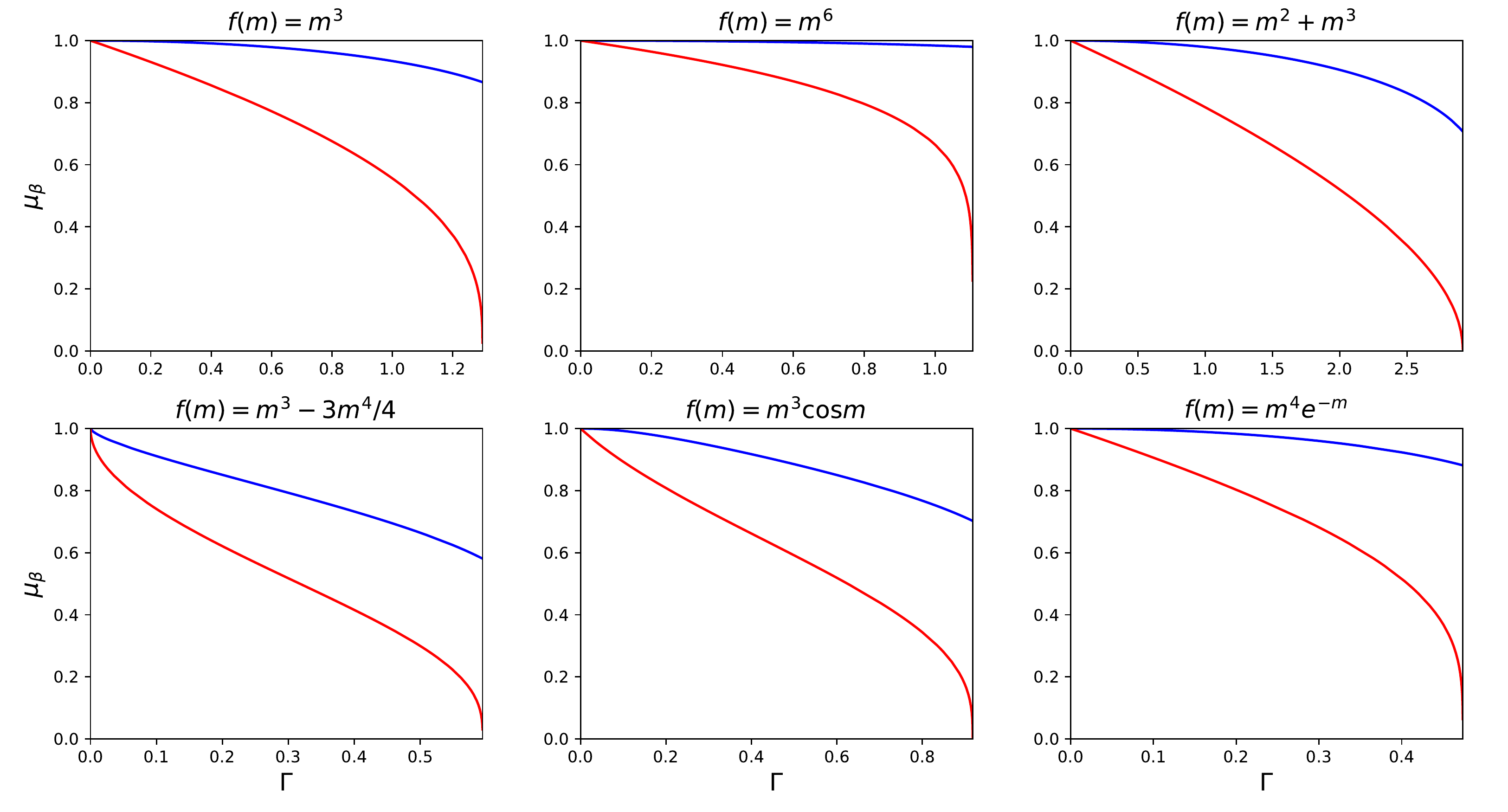}
\caption{Comparison of magnetization (blue, $\beta=2$) against surplus (red, $\beta=1$) for many different fitness functions. In all plots, the right-most value of $\Gamma$ is the transition point $\Gamma_c$.}
\label{fig:mag_collage}
\end{figure*}

For concreteness, we have focused on systems which exhibit a transition from an ordered phase having magnetization $\mu_2 > 0$ to a disordered phase having $\mu_2 = 0$ as $\Gamma$ is increased.
A sufficient condition is that the fitness function $f(m)$ increase monotonically with $m$ and grow no faster than $O(m^2)$ near $m = 0$.
For example, $f(m) = m^2 \textrm{sgn}[m]$ and $f(m) = m^3$ both exhibit such a transition, as shown in Fig.~\ref{fig:potential}.
Note that the former undergoes a continuous transition (in that $\mu$ decreases to 0 continuously) whereas the latter is discontinuous.

The corresponding $\mu_{\beta}$ for these examples are shown in Fig.~\ref{fig:general_mag}.
Considering the upper panel, we see that as $\Gamma \rightarrow \Gamma_c$ from below, the magnetization $\mu_2$ vanishes as $\sqrt{\Gamma_c - \Gamma}$ but the surplus $\mu_1$ vanishes more rapidly as $\Gamma_c - \Gamma$ (the precise scaling can easily be verified analytically).
In the language of critical exponents, the magnetization has exponent $1/2$ whereas the surplus has exponent $1$.

The contrast is even more stark in the lower panel: whereas $\mu_2$ remains finite as $\Gamma \rightarrow \Gamma_c$, $\mu_1$ vanishes.
This behavior is quite generic.
Fig.~\ref{fig:mag_collage} presents the magnetization and surplus for a wide variety of fitness functions, all chosen so that the transition in magnetization is discontinuous.
In all cases, the transition in surplus is nonetheless continuous.

Furthermore, one can prove that the surplus transition is continuous for \textit{any} model which meets our two criteria stated above (namely that $f(m)$ increases monotonically and grows no faster than $O(m^2)$ near $m = 0$).
The proof is given in Appendix~\ref{app:surplus_continuity}.

\subsection{Arbitrary $\beta$}

Our goal is now to understand this phenomenon in more physical terms.
In doing so, it will be convenient to consider the parameter $\beta$ as an arbitrary positive real number.
For reference, recall the expressions
\begin{equation} \label{eq:ground_magnetization_repeat}
\begin{gathered}
\mu_2 = \textrm{argmin}_m \big[ U_-(m) \big] , \quad \epsilon_{\textrm{GS}} = U_-(\mu_2), \\
\kappa(m) \equiv 1 + \frac{U_-(m) - \epsilon_{\textrm{GS}}}{\Gamma \sqrt{1 - m^2}},
\end{gathered}
\end{equation}
from which the exponent of the ground state wavefunction is, for $m \leq \mu_2$ (see Eq.~\eqref{eq:WKB_ground_state_wavefunction}),
\begin{equation} \label{eq:ground_wavefunction_repeat}
\phi(m) = -\frac{1}{2} \int_m^{\mu_2} \textrm{d}m \, \log{\Big( \kappa(m) + \sqrt{\kappa(m)^2 - 1} \Big) },
\end{equation}
and the generalized magnetization $\mu_{\beta}$ is given by
\begin{equation} \label{eq:generalized_surplus_repeat}
\begin{gathered}
\mu_{\beta} = \textrm{argmax}_m \big[ s_{\beta}(m) \big] , \\
s_{\beta}(m) = \left( 1 - \frac{\beta}{2} \right) h(m) + \beta \phi(m),
\end{gathered}
\end{equation}
where $h(m)$, the ``binomial entropy,'' is given by Eq.~\eqref{eq:binomial_entropy}.
We are setting $\phi(\mu_2) = 0$ for convenience.

\begin{figure}[t]
\centering
\includegraphics[width=0.9\columnwidth]{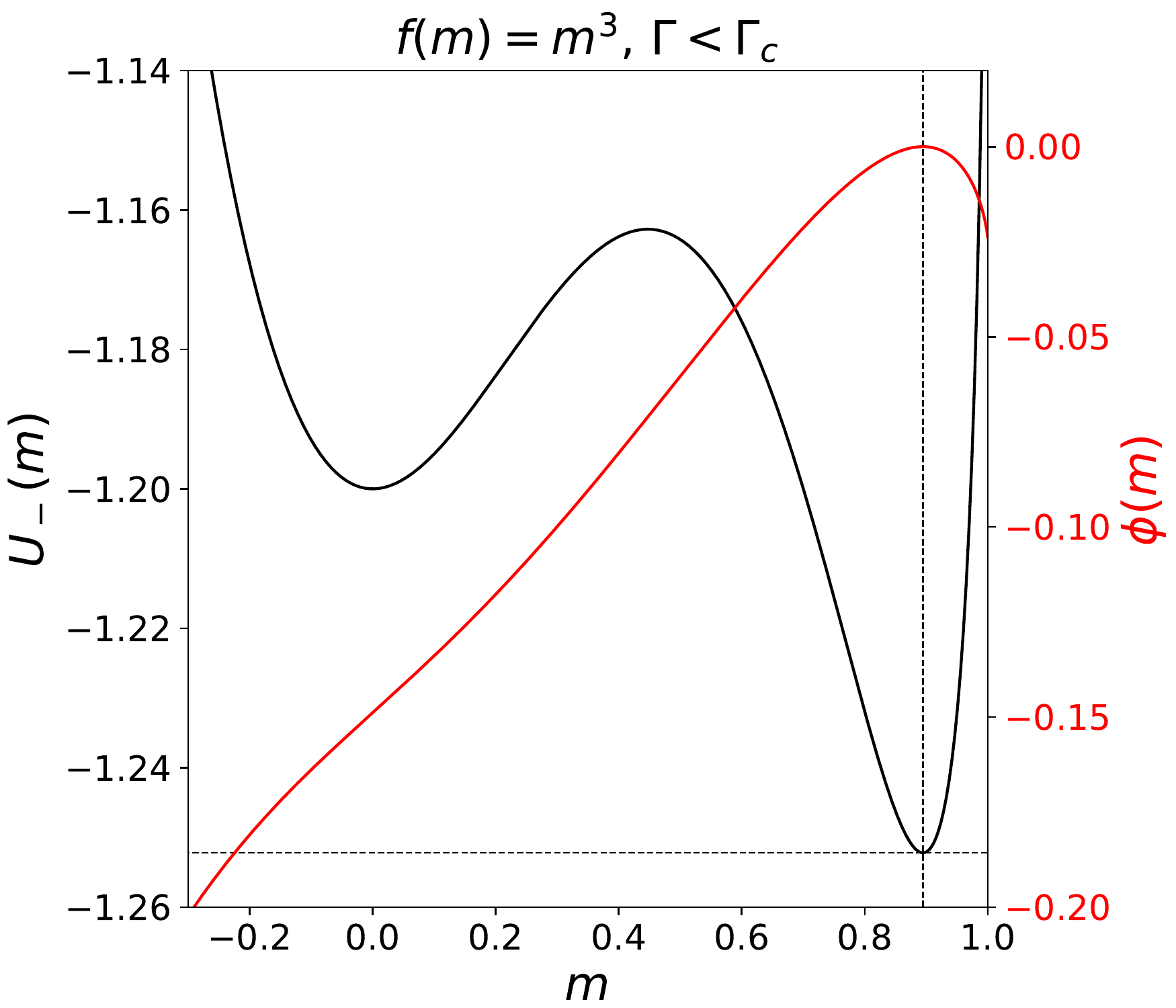}
\includegraphics[width=0.9\columnwidth]{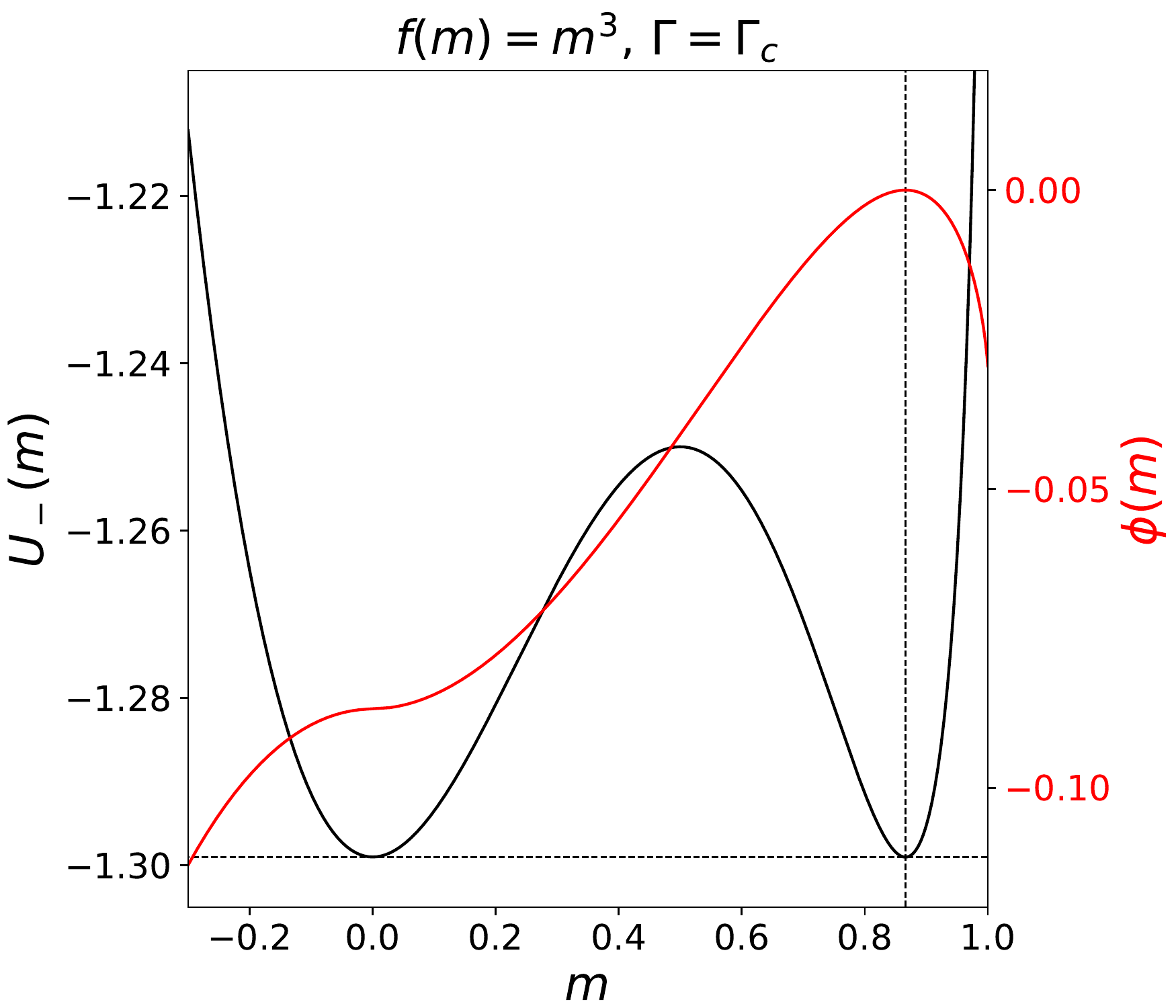}
\caption{Wavefunction $\phi(m)$ for the $f(m) = m^3$ fitness function, both for a value of $\Gamma$ less than $\Gamma_c$ (top) and $\Gamma = \Gamma_c$ (bottom). Wavefunctions are in red, while the corresponding potentials $U_-(m)$ are shown in black, with $\epsilon_{\textrm{GS}}$ and $\mu$ indicated by dashed lines. The precise values of $\Gamma$ are 1.2 (top) and 1.299 (bottom).}
\label{fig:example_wavefunctions}
\end{figure}

Note that the numerator of $\kappa(m) - 1$ is the height of the potential barrier, $U_-(m) - \epsilon_{\textrm{GS}}$.
Furthermore, $\textrm{d} \phi / \textrm{d} m$ increases monotonically with $\kappa(m)$.
Thus the wavefunction behaves roughly as one would find in the WKB treatment of 1D tunneling problems: the slope is zero only at points where the barrier vanishes, and the wavefunction falls off faster in regions where the barrier is larger (albeit with the factor of $\Gamma \sqrt{1 - m^2}$ included).
Figure~\ref{fig:example_wavefunctions} gives an example.
Note that the qualitative features of $\phi(m)$ can be predicted simply from the shape of $U_-(m)$.

\begin{figure}[t]
\centering
\includegraphics[width=0.9\columnwidth]{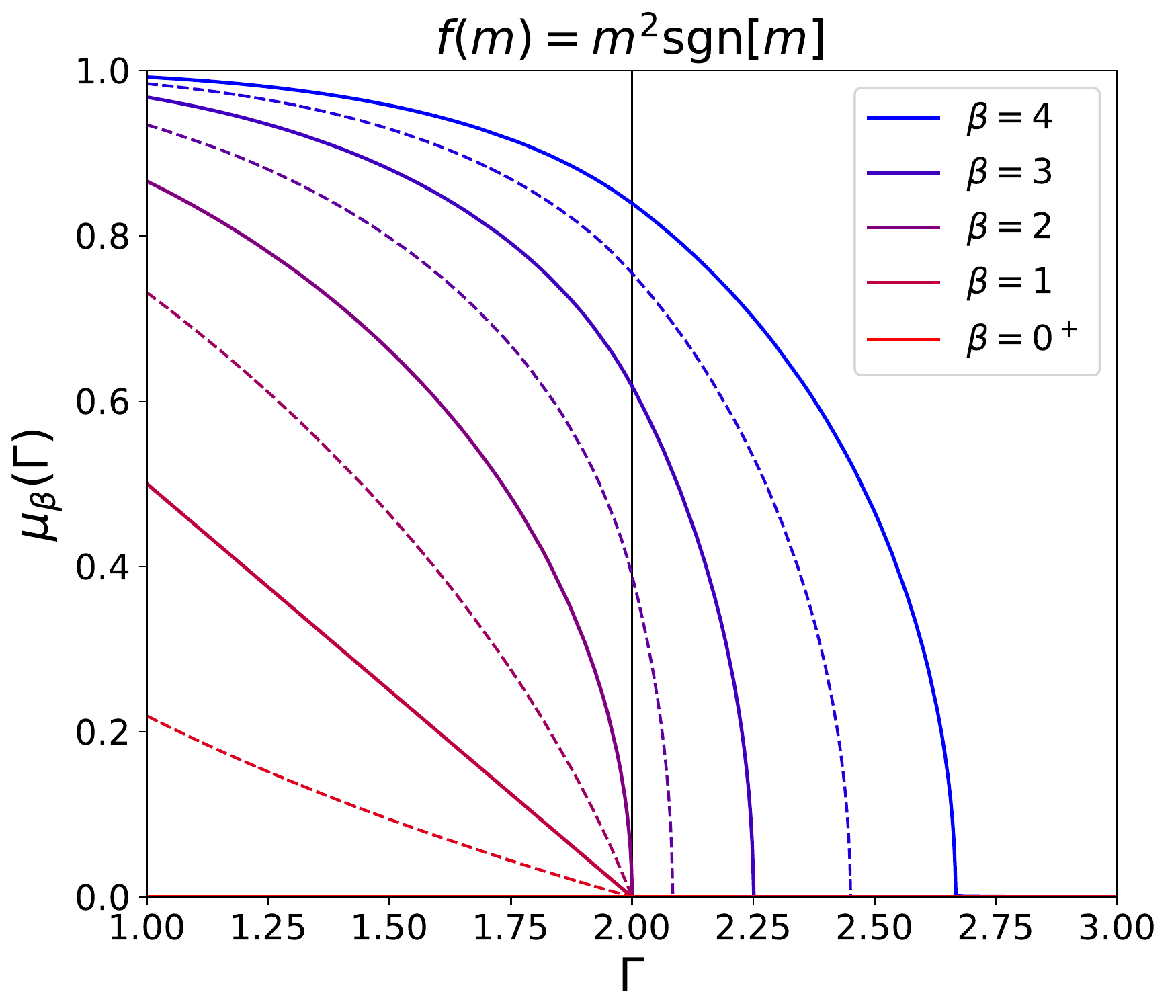}
\includegraphics[width=0.9\columnwidth]{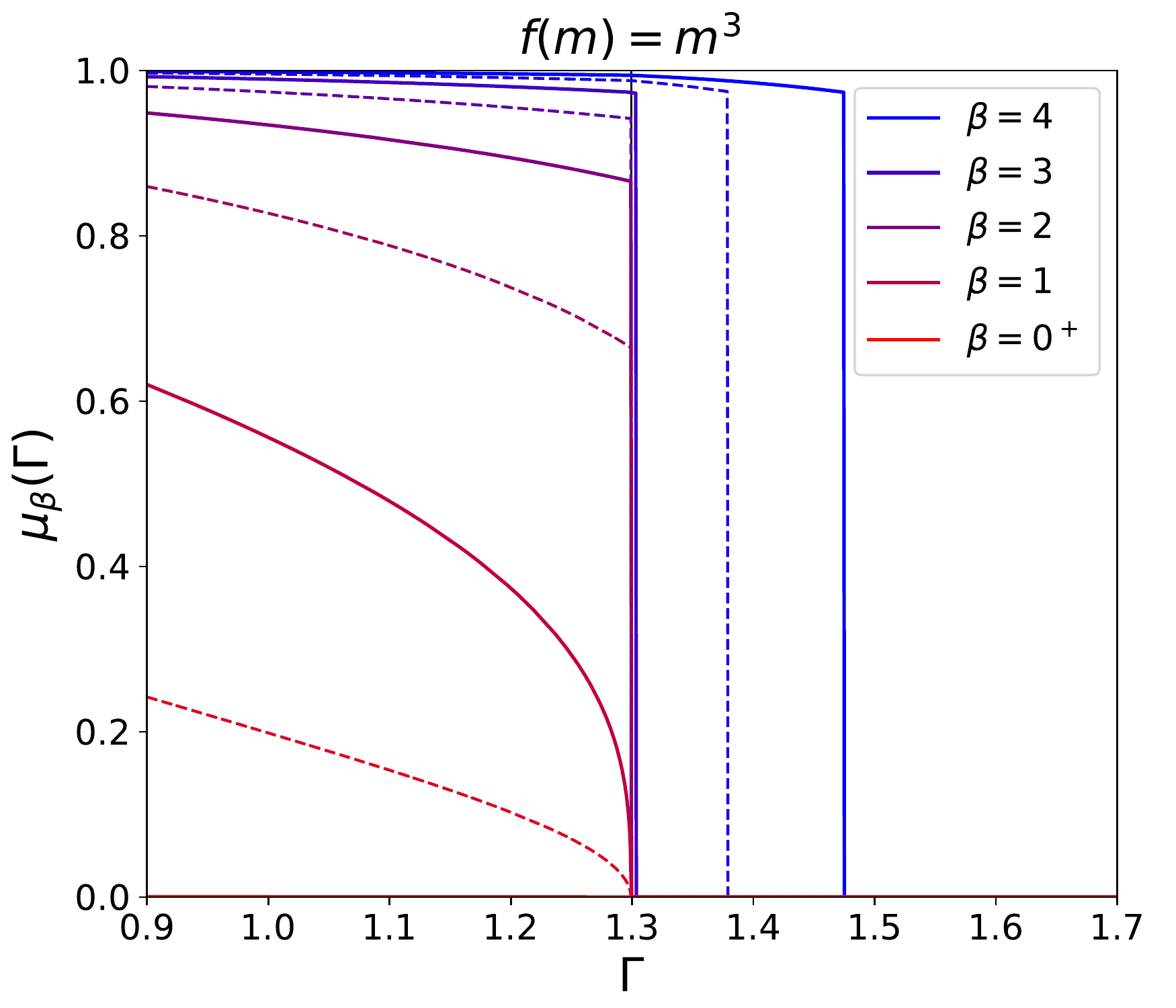}
\caption{Generalized magnetization $\mu_{\beta}$ as a function of $\Gamma$, for various $\beta$ both greater than and less than 2 (decreasing from blue to red). Fitness functions are indicated above each plot. The vertical black lines indicate $\Gamma_c$ for the quantum transition ($\beta = 2$).}
\label{fig:general_mag_extended}
\end{figure}

First consider $\beta < 2$.
A number of results follow immediately from the above discussion:
\begin{itemize}
    \item The surplus is less than the magnetization --- this follows from the fact that $\textrm{d}s_{\beta} / \textrm{d}m |_{m = \mu_2} < 0$ for $\beta < 2$.
    \item The surplus is non-negative --- this follows from $\textrm{d}s_{\beta} / \textrm{d}m > 0$ for $m < 0$.
    \item The surplus is positive for all $\Gamma < \Gamma_c$ --- this follows from $\textrm{d}\phi / \textrm{d}m |_{m = 0} > 0$ (since $U_-(0) > \epsilon_{\textrm{GS}}$) and thus $\textrm{d}s_{\beta} / \textrm{d}m |_{m = 0} > 0$.
    \item The surplus is strictly zero for all $\Gamma > \Gamma_c$ --- both $\textrm{d}\phi / \textrm{d}m$ and $(1 - \beta / 2) h(m)$ are maximized at $m = 0$ when $\Gamma > \Gamma_c$.
\end{itemize}
Note that all these features are borne out in Fig.~\ref{fig:general_mag}.

The maximization of $s_{\beta}(m)$ can be thought of as a competition between two terms.
The binomial contribution $h(m)$ is an entropic term, in that it is maximal at $m = 0$ and strictly concave everywhere.
The wavefunction $\phi(m)$ is an energetic term (although not literally an energy), since it is maximal at $m = \mu_2$.
$\beta$ then plays a role analogous to the inverse temperature in a thermal ensemble: in one limit ($\beta = 0$), the entropic term dominates; in another limit ($\beta = 2$), the energetic term dominates; and for $\beta$ in between, the maximum is at an intermediate value of $m$. 

These considerations together explain why $\mu_{\beta}$ lowers continuously to 0 as $\Gamma \rightarrow \Gamma_c$, at least for small $\beta$.
The wavefunction $\phi(m)$ is a small perturbation to $h(m)$ when $\beta$ is small, and in particular $s_{\beta}(m)$ will be strictly concave for $\beta$ less than a certain non-zero value.
The strict concavity ensures that $\mu_{\beta}$ varies continuously with $\Gamma$, and since we know that $\mu_{\beta} = 0$ at $\Gamma = \Gamma_c$, it follows that $\mu_{\beta} \rightarrow 0$ as $\Gamma \rightarrow \Gamma_c$.

Of course, this argument does not prove that $\mu_1$, the quantity which we are most interested in, must approach 0.
That proof is supplied in Appendix~\ref{app:surplus_continuity}, where we show that $s_1(m)$ cannot be maximized at $m \sim O(1)$ as $\Gamma \rightarrow \Gamma_c$.
In this sense, $\beta = 1$ is sufficiently ``small'' for the above argument to hold.
The critical value of $\beta$ separating continuous from discontinuous $\mu_{\beta}$ can generically be anywhere between 1 and 2, depending on the fitness function.

It is interesting to note that in models with flat fitness functions, such as the single-peak landscape often studied in the literature~\cite{Tarazona1992Error,Franz1997Error}, these conclusions no longer hold.
In particular, one can verify that the surplus of the single-peak landscape ($f(\sigma) = \delta_{m,1}$) is discontinuous: the surplus jumps from 1 to 0 at $\Gamma = 1$.

\begin{figure}[t]
\centering
\includegraphics[width=0.9\columnwidth]{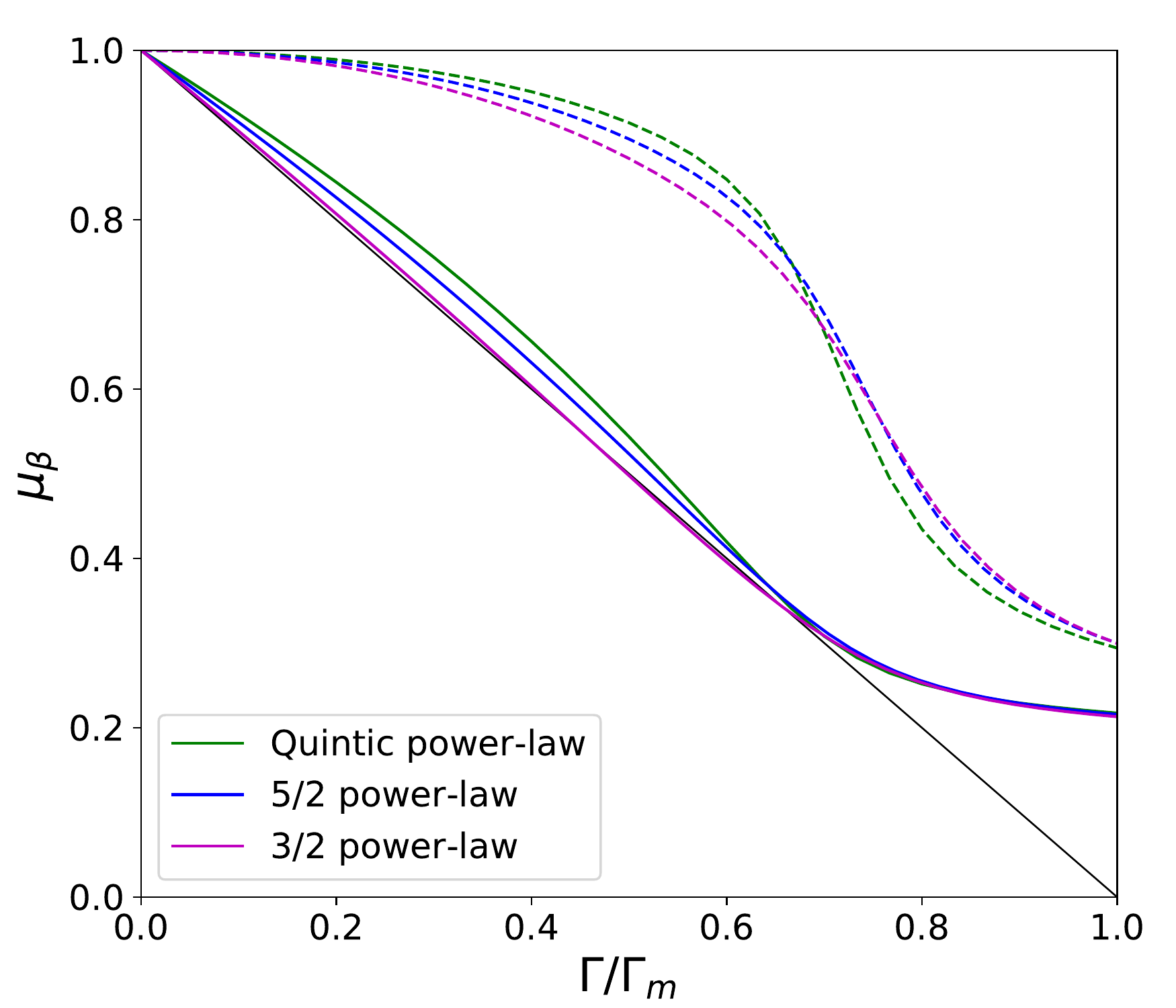}
\caption{Surplus (solid lines) and magnetization (dashed lines) as functions of transverse field, for the Ising models in Eqs.~\eqref{eq:short_power_Hamiltonian},~\eqref{eq:middle_power_Hamiltonian}, and~\eqref{eq:long_power_Hamiltonian}. System size is $N = 22$. For each model, $\Gamma_m$ is the field at which $\mu_2 = 0.3$, chosen simply to normalize the $x$ axis (since the three models have significantly different $\Gamma_c$). The solid black line is merely a straight line drawn for comparison.}
\label{fig:numerics_surplus_comparison}
\end{figure}

Finally, let us briefly consider $\beta > 2$.
The entropy term $(1 - \beta / 2) h(m)$ is now convex, and is minimized at $m = 0$ rather than maximized.
Thus $\mu_{\beta} > \mu_2$.
As a result, $\mu_{\beta}$ need not be zero for all $\Gamma > \Gamma_c$, although it is certainly non-analytic at $\Gamma_c$.
We generically find the behaviors indicated in Fig.~\ref{fig:general_mag_extended}: if the transition in $\mu_2$ is continuous, then the transitions in all $\mu_{\beta}$ will be as well, but at fields increasing with $\beta$.
One can confirm that the critical exponents are the same as for the magnetization, i.e., those of standard mean-field theory.
For discontinuous transitions, the critical field remains at the original $\Gamma_c$ for $\beta$ less than a certain model-dependent value, past which it increases with $\beta$.
This is the situation sketched in Fig.~\ref{fig:phase_diagram}.

\section{Numerical results for short-range models} \label{sec:numerics}

\begin{figure}[t]
\centering
\includegraphics[width=0.9\columnwidth]{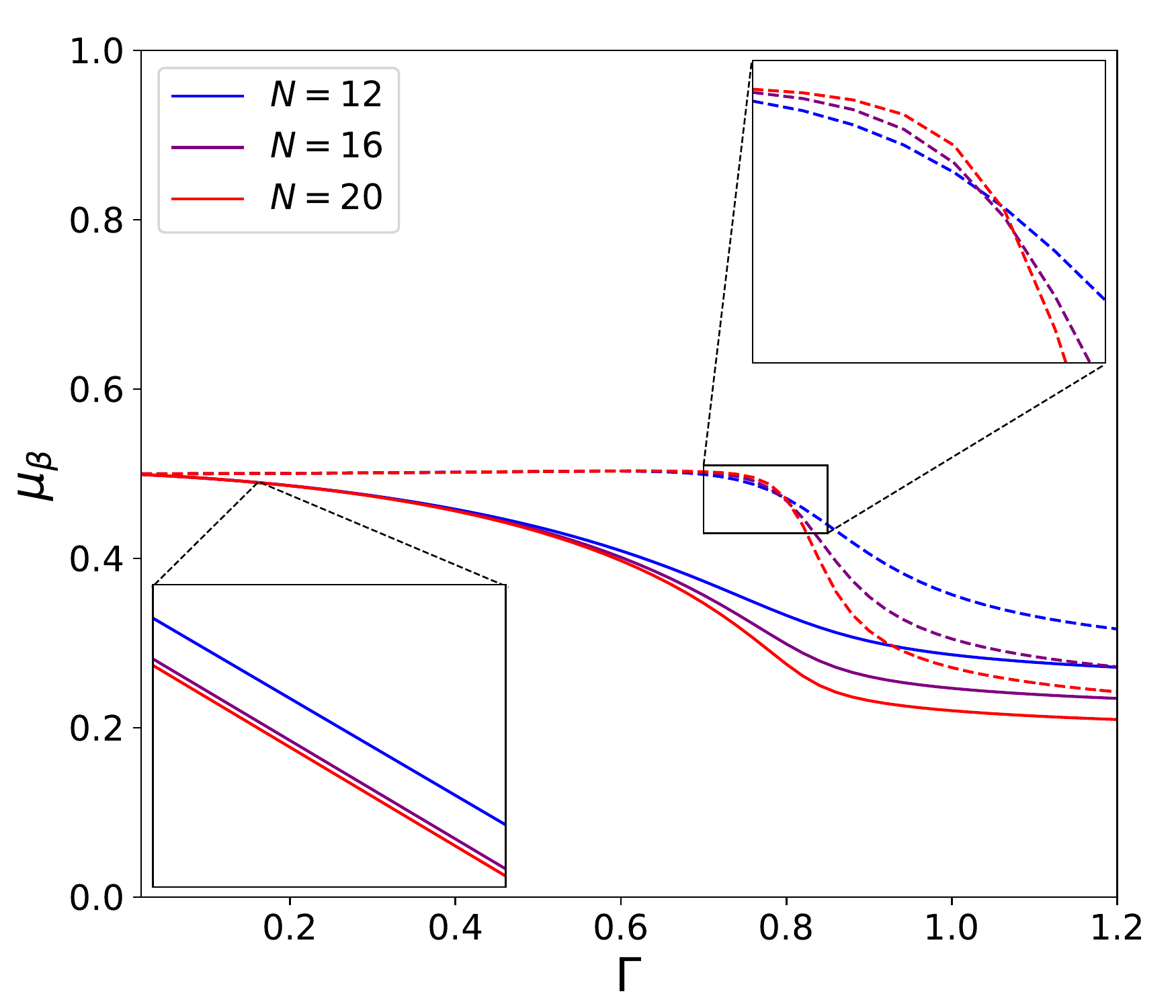}
\caption{Surplus (solid lines) and magnetization (dashed lines) for the Ising model with four-spin interactions, Eq.~\eqref{eq:four_spin_Hamiltonian}. Insets show magnified portions of the plot, demonstrating that the magnetization curves cross each other whereas the surplus curves decrease monotonically with $N$ even at small $\Gamma$.}
\label{fig:numerics_four_spin}
\end{figure}

To reiterate, our analysis of symmetric models identified two important differences between the magnetization and surplus at the critical point: in situations where the magnetization approaches 0 continuously, the surplus is characterized by a different critical exponent ($\mu_1 \sim \Gamma_c - \Gamma$ vs $\mu_2 \sim \sqrt{\Gamma_c - \Gamma}$); and in situations where the magnetization is discontinuous, the surplus is nonetheless continuous.
The rough intuition is that the surplus is more influenced by the entropic effect of there being more configurations having small total spin-$z$ than large.
Yet the analysis used to derive these results relied heavily on the model being symmetric, thus we now investigate whether the conclusions extend to more general systems.

We study a series of one-dimensional transverse-field Ising models through exact diagonalization of the Hamiltonian.
Unfortunately, the accessible system sizes are too small to draw any quantitative conclusions.
One could perform a more systematic study using quantum Monte Carlo -- note that the models considered here do not have sign problems -- together with a finite-size scaling analysis, but we leave that for future work.
The purpose of this section is merely to provide preliminary evidence suggesting that the surplus and magnetization exhibit different critical properties even in non-symmetric models.

One such Hamiltonian, the nearest-neighbor ferromagnetic chain, can be solved analytically as was done in Ref.~\cite{Baake1997Ising}.
The authors showed that the surplus undergoes a non-analyticity at the same $\Gamma_c$ as the magnetization, but with an exponent of $1/2$ rather than the well-known $1/8$ of the magnetization (see also Ref.~\cite{Kaiser1989Surface}).

Here we consider the following fitness functions, all of which are for an $N$-site chain:
\begin{equation} \label{eq:short_power_Hamiltonian}
F_{5} \big( \hat{\sigma}^z \big) = \sum_{i < j} \frac{1}{|i - j|^5} \hat{\sigma}_i^z \hat{\sigma}_j^z,
\end{equation}
\begin{equation} \label{eq:middle_power_Hamiltonian}
F_{5/2} \big( \hat{\sigma}^z \big) = \sum_{i < j} \frac{1}{|i - j|^{5/2}} \hat{\sigma}_i^z \hat{\sigma}_j^z,
\end{equation}
\begin{equation} \label{eq:long_power_Hamiltonian}
F_{3/2} \big( \hat{\sigma}^z \big) = \sum_{i < j} \frac{1}{|i - j|^{3/2}} \hat{\sigma}_i^z \hat{\sigma}_j^z,
\end{equation}
\begin{equation} \label{eq:four_spin_Hamiltonian}
F_{\textrm{Four}} \big( \hat{\sigma}^z \big) = \sum_i \hat{\sigma}_i^z \hat{\sigma}_{i+1}^z - \sum_i \hat{\sigma}_i^z \hat{\sigma}_{i+1}^z \hat{\sigma}_{i+2}^z \hat{\sigma}_{i+3}^z.
\end{equation}
These models do not have analytic solutions, but it is known that the quintic power-law model $F_5$ has the same magnetization exponents as the nearest-neighbor chain, the 3/2 power-law model $F_{3/2}$ has those of mean-field theory, and the 5/2 model $F_{5/2}$ has intermediate exponents~\cite{Dutta2001Phase,Fey2016Critical}.
In general, longer-range interactions have a larger exponent governing the magnetization.
The results shown in Fig.~\ref{fig:numerics_surplus_comparison} are qualitatively consistent with this trend -- the curvature of the curves is smaller for the longer-range models -- and we see that the same trend holds for the surplus.
Again, these observations are hardly quantitative and the differences are quite modest.
One feature which is reasonably clear, however, is that the surplus seems to have a larger exponent than the magnetization in all cases shown.

As for Eq.~\eqref{eq:four_spin_Hamiltonian}, the model with antiferromagnetic four-spin interactions, it has been shown to exhibit a discontinuity in magnetization as one increases $\Gamma$~\cite{DeAlcantara2006Quantum}.
Figure~\ref{fig:numerics_four_spin} shows the surplus and magnetization for $F_{\textrm{Four}}$.
Even though the small system sizes again prohibit quantitative statements, we see that the magnetization curves are consistent with a discontinuous transition: the fall-off near the transition region becomes sharper as system size increases.
The surplus curves do not show any such behavior, and instead are more consistent with a continuous transition.
It thus appears that even in non-symmetric models, the surplus and magnetization transitions can have different orders.

We have not been able to reach any conclusions regarding $\beta > 2$ -- finite size effects are too severe -- but we expect the behavior seen in symmetric models to apply here as well, namely that $\mu_{\beta}$ remains non-zero at $\Gamma_c$ (albeit non-analytically) for sufficiently large $\beta$.
The intuition is again that $\mu_{\beta}$ is shifted relative to $\mu_2$ by an entropic effect, but with the entropic correction now acting to keep $\mu_{\beta} \neq 0$ at $\Gamma_c$.
There are many interesting questions, e.g., the nature of the non-analyticity at $\Gamma_c$ and whether $\mu_{\beta}$ drops to zero at larger fields, and a more systematic study is clearly warranted.

\section{Discussion \& conclusion} \label{sec:discussion}

Many quantum spin Hamiltonians can serve as generators for the evolution of populations under joint mutation and selection, and quantum phase transitions are then associated with error catastrophes.
We have shown here that despite the correspondence between the spin magnetization and the population surplus, the continuity properties of the two can be different.
Transitions in which the magnetization is discontinuous often have a surplus which remains continuous, while continuous transitions come with novel critical exponents for the surplus.

There is a third perspective through which to view these results: the different critical properties of free surfaces as compared to bulk in classical Ising models.
It is well-known that $d$-dimensional quantum Ising systems can be mapped to $(d+1)$-dimensional classical systems, and it is also well-documented that systems with open boundary conditions can have different critical exponents or even orders of transitions at the free surfaces.

To see explicitly that the surplus in mutation-selection models corresponds to a surface magnetization, note that the time evolution of the population (say starting from a specific sequence $\sigma^{(0)}$) can be written compactly as
\begin{equation} \label{eq:population_compact_time_evolution}
\mathcal{N}(\sigma, t) = \langle \sigma | \mathcal{N}(t) \rangle = \langle \sigma | e^{-H t} | \sigma^{(0)} \rangle ,
\end{equation}
which can then be expressed through standard means as the partition function of a classical system.
For example, in the transverse-field models considered here,
\begin{widetext}
\begin{equation} \label{eq:quantum_classical_Trotter_correspondence}
\begin{aligned}
\mathcal{N}(\sigma, t) =& \sum_{\sigma^{(1)} \cdots \sigma^{(M-1)}} \langle \sigma | e^{-H \frac{t}{M}} | \sigma^{(M-1)} \rangle \langle \sigma^{(M-1)} | \cdots | \sigma^{(1)} \rangle \langle \sigma^{(1)} | e^{-H \frac{t}{M}} | \sigma^{(0)} \rangle \\
\sim & \sum_{\sigma^{(1)} \cdots \sigma^{(M-1)}} \exp{\left[ \sum_{m=0}^{M-1} \left( \frac{t}{M} F \big( \sigma^{(m)} \big) + V \big( \sigma^{(m+1)}, \sigma^{(m)} \big) \right) \right] } ,
\end{aligned}
\end{equation}
\end{widetext}
where $V(\sigma, \sigma') \equiv \frac{1}{2} \log{\coth{\frac{\Gamma t}{M}}} \sum_i \sigma_i \sigma'_i$ plus a constant (with $M \rightarrow \infty$ implied).
We see that the transverse field corresponds to a ferromagnetic interaction between $\sigma^{(m)}$ and $\sigma^{(m+1)}$, regardless of the form of the fitness function.

Note that in Eq.~\eqref{eq:quantum_classical_Trotter_correspondence}, $\sigma^{(M)}$ is fixed at $\sigma$.
It is also a ``surface'' layer of spins, in that there is no $\sigma^{(M+1)}$ to interact with.
Finally, to compute the average over the population of any quantity $g(\sigma)$, we evaluate
\begin{equation} \label{eq:quantum_classical_surface_average}
\sum_{\sigma} g(\sigma) \frac{\mathcal{N}(\sigma, t)}{\sum_{\sigma'} \mathcal{N}(\sigma', t)} \propto \sum_{\sigma^{(1)} \cdots \sigma^{(M)}} g \big( \sigma^{(M)} \big) \exp{ \bigg[ \cdots \bigg] },
\end{equation}
where $\cdots$ denotes the exponent in Eq.~\eqref{eq:quantum_classical_Trotter_correspondence}.
For the surplus in particular, we see that it is precisely the average magnetization of the surface layer in the classical Ising model.

This relationship was first discussed in Ref.~\cite{LeuthAusser1987Statistical}, and indeed, many of the previous works comparing surplus to magnetization have been in the language of surface versus bulk magnetization~\cite{Tarazona1992Error,Franz1993Evolutionary}.
In particular, the continuity of magnetization at the surface despite discontinuity in the bulk has been understood as an example of ``wetting''.
The existence of novel surface exponents has also been well-studied in that context~\cite{Binder1972Phase,Lipowsky1984Surface,Binder1990Critical}.

Of course, these considerations alone do not prove that the surplus must behave differently than magnetization at the transition point.
Rather, they simply raise the possibility.
The results we have presented here show that it is indeed a generic phenomenon which occurs in practice.

It is clear that the techniques and ideas of quantum statistical physics can fruitfully be applied to problems in population dynamics.
At the same time, as the above results demonstrate, the population-dynamical analogues of quantum spin systems exhibit novel behaviors which are not simple corollaries to the quantum physics.
The former can also be considered in situations where the latter cannot, such as non-Hermitian models~\cite{Saakian2006Exact}.
Further investigation of quantum systems as population-dynamical models and vice-versa will undoubtedly uncover additional surprises and insights for both fields.

Finally, there is the question of whether the generalized magnetization $\mu_{\beta}$ has physical significance for \textit{arbitrary} $\beta$.
This larger family of observables is useful for understanding the distinction between surplus and magnetization, as can be seen in Fig.~\ref{fig:phase_diagram}, and it would be valuable to know what other information is contained in the $\Gamma$-$\beta$ phase diagram.
There are contexts in which one considers a probability distribution raised to arbitrary powers.
For example, Ref.~\cite{Garrison2018Does} has recently shown that, for certain classes of quantum Hamiltonians, exponentiating the reduced density matrix obtained from an eigenstate at some energy density allows one to probe properties of the system at different energy densities.
Two other situations which come to mind are calculation of Renyi entropies (both classical~\cite{Renyi1961On} and quantum~\cite{MullerLennert2013On}) and multifractality~\cite{Kohmoto1988Entropy,Janssen1998,DeLuca2014Anderson}, and we certainly expect that these are not the only examples.
The implications of our results in these areas is a topic for further study.

\section{Acknowledgements}

The authors would like to thank the Galileo Galilei Institute for Theoretical Physics and the organizers of the workshop ``Breakdown of Ergodicity in Isolated Quantum Systems: From Glassiness to Localization'', where this work was begun.
This research was performed while C.L.B. held a National Institute of Standards and Technology (NIST) National Research Council (NRC) Research
Postdoctoral Associateship Award.
MK is supported by NSF through grant No. DMR-1708280 and SLS by the United States Department  of  Energy  via  grant  No.   DE-SC0016244.

\bibliography{Mutation_Biblio}

\begin{thebibliography}{10}

\bibitem{Eigen1971Selforganization}
M.~Eigen, ``Selforganization of matter and the evolution of biological
  macromolecules,'' {\em Naturwissenschaften}, vol.~58, pp.~465--523, Oct 1971.

\bibitem{Wilke2005Quasispecies}
C.~O. Wilke, ``Quasispecies theory in the context of population genetics,''
  {\em BMC Evolutionary Biology}, vol.~5, no.~1, p.~44, 2005.

\bibitem{Crotty2001RNA}
S.~Crotty, C.~E. Cameron, and R.~Andino, ``Rna virus error catastrophe: Direct
  molecular test by using ribavirin,'' {\em Proceedings of the National Academy
  of Sciences}, vol.~98, p.~6895, 06 2001.

\bibitem{Zhang2003Cytidine}
H.~Zhang, B.~Yang, R.~J. Pomerantz, C.~Zhang, S.~C. Arunachalam, and L.~Gao,
  ``The cytidine deaminase cem15 induces hypermutation in newly synthesized
  hiv-1 dna,'' {\em Nature}, vol.~424, no.~6944, pp.~94--98, 2003.

\bibitem{Anderson2004Viral}
J.~P. Anderson, R.~Daifuku, and L.~A. Loeb, ``Viral error catastrophe by
  mutagenic nucleosides,'' {\em Annual Review of Microbiology}, vol.~58,
  pp.~183--205, 2020/08/18 2004.

\bibitem{GrandePerez2005Suppression}
A.~Grande-P{\'e}rez, E.~L{\'a}zaro, P.~Lowenstein, E.~Domingo, and S.~C.
  Manrubia, ``Suppression of viral infectivity through lethal defection,'' {\em
  Proceedings of the National Academy of Sciences of the United States of
  America}, vol.~102, p.~4448, 03 2005.

\bibitem{Hart2015Error}
G.~R. Hart and A.~L. Ferguson, ``Error catastrophe and phase transition in the
  empirical fitness landscape of hiv,'' {\em Phys. Rev. E}, vol.~91, p.~032705,
  Mar 2015.

\bibitem{Gupta2015Scaling}
V.~Gupta and N.~M. Dixit, ``Scaling law characterizing the dynamics of the
  transition of {HIV}-1 to error catastrophe,'' {\em Physical Biology},
  vol.~12, p.~054001, sep 2015.

\bibitem{Shivam2020Studying}
S.~Shivam, C.~L. Baldwin, J.~Barton, M.~Kardar, and S.~L. Sondhi, ``Studying
  viral populations with tools from quantum spin chains,'' {\em
  arXiv:2003.10668}, 2020.

\bibitem{Note1}
Many works use a slightly different definition of error catastrophe, namely
  when the fraction of wild-type states in the population becomes zero. We use
  the definition involving average surplus because it is more natural from a
  statistical-physics perspective. See as well Ref.~\cite {Franz1997Error}.

\bibitem{Baake1997Ising}
E.~Baake, M.~Baake, and H.~Wagner, ``Ising quantum chain is equivalent to a
  model of biological evolution,'' {\em Phys. Rev. Lett.}, vol.~78,
  pp.~559--562, Jan 1997.

\bibitem{Wagner1998Ising}
H.~Wagner, E.~Baake, and T.~Gerisch, ``Ising quantum chain and sequence
  evolution,'' {\em Journal of Statistical Physics}, vol.~92, pp.~1017--1052,
  Sep 1998.

\bibitem{Tarazona1992Error}
P.~Tarazona, ``Error thresholds for molecular quasispecies as phase
  transitions: From simple landscapes to spin-glass models,'' {\em Phys. Rev.
  A}, vol.~45, pp.~6038--6050, Apr 1992.

\bibitem{Hermisson2001Four}
J.~Hermisson, H.~Wagner, and M.~Baake, ``Four-state quantum chain as a model of
  sequence evolution,'' {\em Journal of Statistical Physics}, vol.~102, no.~1,
  pp.~315--343, 2001.

\bibitem{Franz1997Error}
S.~Franz and L.~Peliti, ``Error threshold in simple landscapes,'' {\em Journal
  of Physics A: Mathematical and General}, vol.~30, pp.~4481--4487, jul 1997.

\bibitem{Baake1998Quantum}
E.~Baake, M.~Baake, and H.~Wagner, ``Quantum mechanics versus classical
  probability in biological evolution,'' {\em Phys. Rev. E}, vol.~57,
  pp.~1191--1192, Jan 1998.

\bibitem{Hermisson2002Mutation}
J.~Hermisson, O.~Redner, H.~Wagner, and E.~Baake, ``Mutation--selection
  balance: Ancestry, load, and maximum principle,'' {\em Theoretical Population
  Biology}, vol.~62, no.~1, pp.~9--46, 2002.

\bibitem{Peliti2002Quasispecies}
L.~Peliti, ``Quasispecies evolution in general mean-field landscapes,'' {\em
  Europhysics Letters ({EPL})}, vol.~57, pp.~745--751, mar 2002.

\bibitem{Saakian2004Solvable}
D.~B. Saakian, C.-K. Hu, and H.~Khachatryan, ``Solvable biological evolution
  models with general fitness functions and multiple mutations in parallel
  mutation-selection scheme,'' {\em Phys. Rev. E}, vol.~70, p.~041908, Oct
  2004.

\bibitem{Bapst2012On}
V.~Bapst and G.~Semerjian, ``On quantum mean-field models and their quantum
  annealing,'' {\em Journal of Statistical Mechanics: Theory and Experiment},
  vol.~2012, no.~06, p.~P06007, 2012.

\bibitem{Zhao2014Three}
B.~Zhao, M.~C. Kerridge, and D.~A. Huse, ``Three species of schr\"odinger cat
  states in an infinite-range spin model,'' {\em Phys. Rev. E}, vol.~90,
  p.~022104, Aug 2014.

\bibitem{Garg1998Application}
A.~Garg, ``Application of the discrete wentzel--kramers--brillouin method to
  spin tunneling,'' {\em Journal of Mathematical Physics}, vol.~39, no.~10,
  pp.~5166--5179, 1998.

\bibitem{Saakian2007New}
D.~B. Saakian, ``A new method for the solution of models of biological
  evolution: Derivation of exact steady-state distributions,'' {\em Journal of
  Statistical Physics}, vol.~128, no.~3, pp.~781--798, 2007.

\bibitem{Note2}
This is analogous to the situation in single-particle bound state problems,
  where states must have energies greater than the minimum of the potential in
  order to be normalizable.

\bibitem{Kaiser1989Surface}
C.~Kaiser and I.~Peschel, ``Surface and corner magnetizations in the
  two-dimensional ising model,'' {\em Journal of Statistical Physics}, vol.~54,
  no.~3, pp.~567--579, 1989.

\bibitem{Dutta2001Phase}
A.~Dutta and J.~K. Bhattacharjee, ``Phase transitions in the quantum ising and
  rotor models with a long-range interaction,'' {\em Phys. Rev. B}, vol.~64,
  p.~184106, Oct 2001.

\bibitem{Fey2016Critical}
S.~Fey and K.~P. Schmidt, ``Critical behavior of quantum magnets with
  long-range interactions in the thermodynamic limit,'' {\em Phys. Rev. B},
  vol.~94, p.~075156, Aug 2016.

\bibitem{DeAlcantara2006Quantum}
O.~F. de~Alcantara~Bonfim and J.~Florencio, ``Quantum phase transitions in the
  transverse one-dimensional ising model with four-spin interactions,'' {\em
  Phys. Rev. B}, vol.~74, p.~134413, Oct 2006.

\bibitem{LeuthAusser1987Statistical}
I.~Leuth{\"A}usser, ``Statistical mechanics of eigen's evolution model,'' {\em
  Journal of Statistical Physics}, vol.~48, no.~1, pp.~343--360, 1987.

\bibitem{Franz1993Evolutionary}
S.~Franz, L.~Peliti, and M.~Sellitto, ``An evolutionary version of the random
  energy model,'' {\em Journal of Physics A: Mathematical and General},
  vol.~26, pp.~L1195--L1199, dec 1993.

\bibitem{Binder1972Phase}
K.~Binder and P.~C. Hohenberg, ``Phase transitions and static spin correlations
  in ising models with free surfaces,'' {\em Phys. Rev. B}, vol.~6,
  pp.~3461--3487, Nov 1972.

\bibitem{Lipowsky1984Surface}
R.~Lipowsky, ``Surface‐induced order and disorder: Critical phenomena at
  first‐order phase transitions (invited),'' {\em Journal of Applied
  Physics}, vol.~55, pp.~2485--2490, 2020/01/06 1984.

\bibitem{Binder1990Critical}
K.~Binder and D.~P. Landau, ``Critical phenomena at surfaces,'' {\em Physica A:
  Statistical Mechanics and its Applications}, vol.~163, no.~1, pp.~17--30,
  1990.

\bibitem{Saakian2006Exact}
D.~B. Saakian and C.-K. Hu, ``Exact solution of the eigen model with general
  fitness functions and degradation rates,'' {\em Proceedings of the National
  Academy of Sciences of the United States of America}, vol.~103, p.~4935, 03
  2006.

\bibitem{Garrison2018Does}
J.~R. Garrison and T.~Grover, ``Does a single eigenstate encode the full
  hamiltonian?,'' {\em Phys. Rev. X}, vol.~8, p.~021026, Apr 2018.

\bibitem{Renyi1961On}
A.~Renyi, ``On measures of entropy and information,'' pp.~547--561, 1961.

\bibitem{MullerLennert2013On}
M.~M{\"u}ller-Lennert, F.~Dupuis, O.~Szehr, S.~Fehr, and M.~Tomamichel, ``On
  quantum r{\'e}nyi entropies: A new generalization and some properties,'' {\em
  Journal of Mathematical Physics}, vol.~54, p.~122203, 2020/08/31 2013.

\bibitem{Kohmoto1988Entropy}
M.~Kohmoto, ``Entropy function for multifractals,'' {\em Phys. Rev. A},
  vol.~37, pp.~1345--1350, Feb 1988.

\bibitem{Janssen1998}
M.~Janssen, ``Statistics and scaling in disordered mesoscopic electron
  systems,'' {\em Physics Reports}, vol.~295, no.~1, pp.~1 -- 91, 1998.

\bibitem{DeLuca2014Anderson}
A.~De~Luca, B.~L. Altshuler, V.~E. Kravtsov, and A.~Scardicchio, ``Anderson
  localization on the bethe lattice: Nonergodicity of extended states,'' {\em
  Phys. Rev. Lett.}, vol.~113, p.~046806, Jul 2014.

\end{thebibliography}
\bibliographystyle{ieeetr}

\appendix

\section{Boundary conditions} \label{app:boundary_conditions}

In the main text, we derived Eq.~\eqref{eq:WKB_zeroth_order_equation}, written here as
\begin{equation} \label{eq:leading_order_exponent_equation_v1}
\cosh{\left( 2 \frac{\textrm{d}\phi}{\textrm{d}m} \right) } = \kappa(m), \quad \kappa(m) \equiv \frac{-\epsilon - f(m)}{\Gamma \sqrt{1 - m^2}}.
\end{equation}
At every $m$, this equation has two solutions:
\begin{equation} \label{eq:leading_order_exponent_equation_v2}
\frac{\textrm{d}\phi}{\textrm{d}m} = \frac{1}{2} \log{\Big( \kappa(m) \pm \sqrt{\kappa(m)^2 - 1} \Big) }.
\end{equation}
Just as in one-dimensional tunneling problems, the boundary conditions determine which sign to use.
Here, we show that the correct sign is $+$ near $m = -1$ and $-$ near $m = 1$.
This then fixes the allowed values of $\epsilon$, as discussed in the main text.

Starting from the Schrodinger equation, Eq.~\eqref{eq:symmetric_Schrodinger_equation}, first set $M = N - 2J$ with $J \ll O(N)$.
To leading order in $J/N$, the equation simplifies to
\begin{equation} \label{eq:symmetric_schrodinger_right_endpoint}
\Psi(J+1) = -\frac{E + Nf(1)}{\Gamma \sqrt{N(J+1)}} \Psi(J) - \sqrt{\frac{J}{J+1}} \Psi(J-1).
\end{equation}
The second term on the right-hand side will turn out to be subleading compared to the first, and so we omit it.
Defining $\Phi(J) \equiv \log{\Psi(J)}$, we have
\begin{equation} \label{eq:right_endpoint_log_equation}
\Phi(J+1) = \Phi(J) + \frac{1}{2} \log{\frac{N}{J+1}} + \log{\frac{-\epsilon - f(1)}{\Gamma}},
\end{equation}
which can be easily solved:
\begin{equation} \label{eq:right_endpoint_solution}
\Phi(J) = \Phi(0) + J \log{\frac{-\epsilon - f(1)}{\Gamma}} + \frac{1}{2} \sum_{K=1}^J \log{\frac{N}{K}}.
\end{equation}
This is an exact expression for the solution of the Schrodinger equation, which does not rely on taking any continuum limit.

Let us now compare Eq.~\eqref{eq:right_endpoint_solution} to what we would find expanding the continuum Eq.~\eqref{eq:leading_order_exponent_equation_v2} in $1 - m$.
Note that $\kappa(m) \rightarrow \infty$ as $m \rightarrow 1$ (at least for $\epsilon \neq f(1)$).
Thus
\begin{equation} \label{eq:continuum_equation_right_endpoint}
\begin{aligned}
\frac{\textrm{d}\phi}{\textrm{d}m} \sim & \pm \frac{1}{2} \log{2\kappa(m)} \\
\sim & \pm \frac{1}{2} \log{\frac{\sqrt{2} \big( -\epsilon - f(1) \big)}{\Gamma}} \mp \frac{1}{4} \log{(1 - m)}.
\end{aligned}
\end{equation}
Integrating from $m = 1$ to $m = 1 - 2j$ gives
\begin{equation} \label{eq:continuum_right_endpoint_solution}
\phi(1 - 2j) = \phi(1) \mp j \log{\frac{-\epsilon - f(1)}{\Gamma}} \mp \frac{1}{2} \int_0^j \textrm{d}k \log{\frac{1}{k}}.
\end{equation}
Comparing Eqs.~\eqref{eq:right_endpoint_solution} and~\eqref{eq:continuum_right_endpoint_solution} (noting that $\Phi(J) = N \phi(1 - 2j)$ by definition), we see that the lower sign is needed for the continuum result to agree with the exact expression.

A similar analysis holds near $m = -1$.
Writing $M = -N + 2J$ and $\Phi(J) = N \phi(-1 + 2j)$, we find
\begin{equation} \label{eq:left_endpoint_solution}
\Phi(J) = \Phi(0) + J \log{\frac{-\epsilon - f(-1)}{\Gamma}} + \frac{1}{2} \sum_{K=1}^J \log{\frac{N}{K}},
\end{equation}
to be compared with
\begin{equation} \label{eq:continuum_left_endpoint_solution}
\phi(-1 + 2j) = \phi(-1) \pm j \log{\frac{-\epsilon - f(-1)}{\Gamma}} \pm \frac{1}{2} \int_0^j \textrm{d}k \log{\frac{1}{k}}.
\end{equation}
The upper sign is needed for the two expressions to agree.

Thus a valid solution to the Schrodinger equation must indeed obey Eq.~\eqref{eq:leading_order_exponent_equation_v2} with the plus sign near $m = -1$ and the minus sign near $m = 1$.

\section{Continuity of the surplus} \label{app:surplus_continuity}

Here we show that the surplus must approach 0 continuously as $\Gamma \rightarrow \Gamma_c$, for any symmetric model which meets the criteria given in the main text ($f(m)$ increasing monotonically with $m$ and growing no faster than $O(m^2)$ near $m = 0$).
We do so by proving that at $\Gamma_c$, $\textrm{d} s_1 / \textrm{d}m \leq 0$ for all $m \geq 0$, with equality only at $m = 0$.
Since $s_1(m)$ varies continuously as $\Gamma$ approaches $\Gamma_c$ from below (it is only when the argmin of $U_-(m)$ jumps as $\Gamma$ \textit{crosses} $\Gamma_c$ that there is a non-analyticity), this implies that for $\Gamma$ infinitesimally less than $\Gamma_c$, the argmax of $s_1(m)$ cannot be at any non-infinitesimal $m$, i.e., $\mu_1$ is continuous in $\Gamma$.

Without loss of generality, we can take $f(0) = 0$.
At $\Gamma_c$, which is the field strength at which $U_-(0) = \epsilon_{\textrm{GS}}$, we thus have $\epsilon_{\textrm{GS}} = -\Gamma_c$.
Then
\begin{equation} \label{eq:critical_kappa_function}
\kappa(m) = \frac{\Gamma_c - f(m)}{\Gamma_c \sqrt{1 - m^2}}.
\end{equation}
We thus write $\textrm{d}s_1 / \textrm{d}m$ as
\begin{widetext}
\begin{equation} \label{eq:critical_surplus_action_expression}
\begin{aligned}
\frac{\textrm{d}s_1}{\textrm{d}m} =& \, \frac{1}{4} \log{\frac{1 - m}{1 + m}} + \frac{1}{2} \log{\Big( \kappa(m) + \sqrt{\kappa(m)^2 - 1} \Big) } \\
=& \, \frac{1}{2} \log{\frac{1}{1 + m}} + \frac{1}{2} \log{\left( 1 - \frac{f(m)}{\Gamma_c} + \sqrt{m^2 - 2 \frac{f(m)}{\Gamma_c} + \frac{f(m)^2}{\Gamma_c^2}} \right) }.
\end{aligned}
\end{equation}

Since the minimum of $U_-(m)$ is not at $m = 1$, we know that
\begin{equation} \label{eq:critical_potential_endpoint_bound}
U_-(1) = -f(1) > \epsilon_{\textrm{GS}} = -\Gamma_c,
\end{equation}
and since $f(m)$ is monotonic in $m$, it follows that for all $m \in [0, 1]$, 
\begin{equation} \label{eq:critical_potential_bounds}
0 \leq f(m) < \Gamma_c.
\end{equation}
We thus have the following chain of inequalities:
\begin{equation} \label{eq:critical_action_inequalities}
\begin{aligned}
-2 \frac{f(m)}{\Gamma_c} + \frac{f(m)^2}{\Gamma_c^2} \leq & \; 0 \\
\Rightarrow -2 \frac{f(m)}{\Gamma_c} + \frac{f(m)^2}{\Gamma_c^2} \leq & \; m^2 \left( -2 \frac{f(m)}{\Gamma_c} + \frac{f(m)^2}{\Gamma_c^2} \right) \\
\Rightarrow m^2 - 2 \frac{f(m)}{\Gamma_c} + \frac{f(m)^2}{\Gamma_c^2} \leq & \; m^2 \left( 1 - \frac{f(m)}{\Gamma_c} \right) ^2 \\
\Rightarrow 1 - \frac{f(m)}{\Gamma_c} + \sqrt{m^2 - 2 \frac{f(m)}{\Gamma_c} + \frac{f(m)^2}{\Gamma_c^2}} \leq & \; \big( 1 + m \big) \left( 1 - \frac{f(m)}{\Gamma_c} \right) .
\end{aligned}
\end{equation}
\end{widetext}

Inserting into Eq.~\eqref{eq:critical_surplus_action_expression}, we have simply
\begin{equation} \label{eq:critical_action_final_bound}
\frac{\textrm{d}s_1}{\textrm{d}m} \leq \frac{1}{2} \log{\left( 1 - \frac{f(m)}{\Gamma_c} \right) }.
\end{equation}
Since $f(m)$ is monotonic and $f(0) = 0$, this establishes what we claimed: $\textrm{d}s_1 / \textrm{d}m \leq 0$ with equality only at $m = 0$.

In fact, $s_1(m)$ has nice properties which allow us to determine the surplus quite simply.
Starting from the upper line of Eq.~\eqref{eq:critical_surplus_action_expression} and setting $\textrm{d}s_1 / \textrm{d}m = 0$, we have
\begin{equation} \label{eq:surplus_simple_evaluation_start}
\kappa(m) + \sqrt{\kappa(m)^2 - 1} = \sqrt{\frac{1 + m}{1 - m}}.
\end{equation}
Using the explicit expression for $\kappa(m)$ (note that here we are considering arbitrary $\Gamma$), this becomes
\begin{equation} \label{eq:surplus_simple_evaluation_explicit}
-\epsilon - f(m) + \sqrt{\big( \epsilon + f(m) \big) ^2 - \Gamma^2 (1 - m^2)} = \Gamma (1 + m),
\end{equation}
which can be simplified considerably to
\begin{equation} \label{eq:surplus_simple_evaluation_solution}
f(m) = -\epsilon - \Gamma.
\end{equation}
The surplus is given merely by the solution to Eq.~\eqref{eq:surplus_simple_evaluation_solution}.

This result holds for all $\Gamma$, and thus is quite useful in of itself.
Furthermore, it gives an immediate alternate proof that the surplus is continuous at $\Gamma_c$ (albeit one that does not generalize to other values of $\beta$): since $\epsilon$ approaches $-\Gamma_c$ continuously as $\Gamma \rightarrow \Gamma_c$, the solution of Eq.~\eqref{eq:surplus_simple_evaluation_solution} for any monotonic $f(m)$ must approach 0 continuously.

\end{document}